\documentclass[lettersize,journal]{IEEEtran}
\usepackage{array}
\usepackage[caption=false,font=normalsize,labelfont=sf,textfont=sf]{subfig}
\usepackage{textcomp}
\usepackage{stfloats}
\usepackage{url}
\usepackage{verbatim}
\usepackage{graphicx}
\usepackage{cite}
\usepackage{xcolor}
\usepackage{booktabs} 
\usepackage{threeparttable}
\usepackage{multicol}
\usepackage{multirow}
\usepackage{enumitem}
\usepackage[cmtip,all]{xy}
\usepackage{listings}
\usepackage{tikz}
\usetikzlibrary{shapes.geometric}
\usepackage{makecell}
\usepackage{pifont}
\usepackage{cuted}
\usepackage{titlesec}
\usepackage{cuted}

\usepackage{hyperref}
\hypersetup{hidelinks}

\newcommand{\orcid}[1]{\href{https://orcid.org/#1}{\includegraphics[width=8pt]{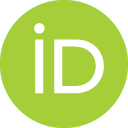}}}

\usepackage[ruled]{algorithm2e}
\usepackage{lipsum}
\usepackage{xurl}

\usepackage{amsmath,amsfonts,amsthm,amssymb}
\usepackage{mathptmx}
\usepackage[mathscr]{eucal}
\DeclareMathAlphabet{\mathbfcal}{OMS}{cmsy}{b}{n}

\usepackage[nogroupskip, numberedsection, acronym]{glossaries}
\makenoidxglossaries

\newacronym{mpc}{MPC}{Multi-Party Computation}
\newacronym{put}{PUT}{Public Transaction}
\newacronym{prt}{PRT}{Private Transaction}
\newacronym{mpt}{MPT}{Multi-Party Transaction}
\newacronym{mpp}{MPP}{Multi-Party Program}
\newacronym{zkp}{ZKP}{Zero-Knowledge Proof}
\newacronym{he}{HE}{Homomorphic Encryption}
\newacronym{tee}{TEE}{Trusted Execution Environment}
\newacronym{dpc}{DPC}{Decentralized Private Computation}
\newacronym{pop}{PoP}{Proof of Publication}
\newacronym{cca2}{IND-CCA2}{Indistinguishability under Adaptive CCA}
\newacronym{cma}{EUF-CMA}{Existential Unforgeability under CMA}
\newacronym{ttp}{TTP}{Trusted Third Party}

\usepackage{xcolor}
\definecolor{pink}{RGB}{255,160,122}
\definecolor{main}{RGB}{221, 157, 118} 

\newcommand{\codename}[0]{\textsc{DeCloak}\xspace}
\newcommand{\myparagraph}[1]{\noindent{\textbf{#1.}}}

\newcommand{\cf}{\hbox{\emph{cf.}}\xspace}

\newcommand{\etal}{\hbox{\emph{et al.}}\xspace}

\newcommand{\eg}{\hbox{\emph{e.g.}}\xspace}

\newcommand{\ie}{\hbox{\emph{i.e.}}\xspace}

\newcommand{\longsquiggly}{\xymatrix{{}\ar@{~>}[r]&{}}}

\newcommand{\code}[1]{{\texttt{\footnotesize #1}}}

\usepackage{tcolorbox}
\tcbuselibrary{skins}

\newtcolorbox{theorem-box}
{
  enhanced jigsaw,
  drop shadow=black!50!white,
  colback=gray!30
}

\newtcolorbox{definition-box}
{
  drop shadow=black!50!white,
  colback=white
}

\newtcolorbox{lemma-box}
{
  drop shadow=black!50!white,
  colback=gray!10
}

\usepackage[english]{babel}
\usepackage{amsthm}
\theoremstyle{theorem}
\newtheorem{informal-theorem}{Theorem}
\newtheorem{theorem}{Theorem}
\newtheorem{definition}{Definition}
\newtheorem{lemma}[theorem]{Lemma}

\begin{document}

\title{\textsc{\codename}: Enable Secure and Cheap \acrlong{mpt}s on Legacy Blockchains by a Minimally Trusted \acrshort{tee} Network}

\author{Qian~Ren\orcid{0000-0002-2617-1321},
        Yue~Li\orcid{0000-0002-4137-619X},
        Yingjun~Wu\orcid{0000-0001-6567-7838},
        Yuchen~Wu\orcid{0009-0000-0231-4570},
        Hong~Lei\orcid{0000-0002-6564-1568},
        Lei~Wang\orcid{0000-0001-7170-2825},
        and~Bangdao~Chen\orcid{0000-0003-3225-4286}
\thanks{This work was supported in part by the National Key R\&D Program of China (No. 2021YFB2700600); in part by the Finance Science and Technology Project of Hainan Province (No. ZDKJ2020009); in part by the National Natural Science Foundation of China (No. 62163011); in part by the Research Startup Fund of Hainan University under Grant KYQD(ZR)-21071. (\textit{Corresponding author: Hong Lei})\\
\indent Q. Ren is with the The Blockhouse Technology Ltd., Oxford OX2 6XJ, UK. (email: qianren1024@gmail).\\
\indent Y. Li is with the Peking University, Beijing, China 100871. (email: liyue\_cs@pku.edu.cn) \\
\indent L. Wang is with the Department of Computer Science and Engineering, Shanghai Jiao Tong University, Shanghai, China 200240. (email: wanglei@cs.sjtu.edu.cn) \\ 
\indent H. Lei is with School of Cyberspace Security (School of Cryptology), Hainan University, Hainan, China 570228. \\
\indent Yi. Wu, Yu. Wu, H. Lei, and B. Chen are with the Oxford-Hainan Blockchain Research Institute and SSC Holding Company Ltd., Wok Park, Laocheng, Chengmai, Hainan, 571924 China. (email: {yingjun, yuchen, leihong, bangdao}@oxhainan.org). 
}
}




\maketitle

\begin{abstract}
As the confidentiality and scalability of smart contracts have become a crucial demand of blockchains, off-chain contract execution frameworks have been promising. Some have recently expanded off-chain contracts to \acrfull{mpc}, which seek to transition the on-chain states by off-chain \acrshort{mpc}. The most general problem among these solutions is \acrshort{mpt}, since its off-chain \acrshort{mpc} takes on- and off-chain inputs, delivers on- and off-chain outputs, and can be publicly verified by the blockchain, thus capable of covering more scenarios. However, existing \acrfull{mpt} solutions lack at least one of data availability, financial fairness, delivery fairness, and delivery atomicity. The data availability means entities can independently access the data required to rebuild new states and verify outputs; financial fairness implies at least one adversary will be punished monetarily; delivery fairness means parties can receive their outputs at almost the same time; delivery atomicity means that parties receive their outputs and new states are committed must both happen or neither. These properties are crucially valued by communities, \eg, the Ethereum community and users. Even worse, these solutions require high-cost interactions between the blockchain and off-chain systems. 

This paper proposes a novel \acrshort{mpt}-enabled off-chain contract execution framework, \codename. \codename is the first to achieve data availability of \acrshort{mpt}, and our method can apply to other fields that seek to persist user data on-chain. Moreover, \codename solves all mentioned shortcomings with lower gas cost and weaker assumption. Specifically, \codename tolerates all-but-one Byzantine party and \acrshort{tee} executors. Evaluating on 10 \acrshort{mpt}s, \codename reduces the gas cost of the SOTA, Cloak, by 65.6\%. Consequently, we are the first to not only achieve such level secure \acrshort{mpt} in practical assumption, but also demonstrate that evaluating \acrshort{mpt} in the comparable average gas cost to Ethereum transactions is possible. And the cost superiority of \codename increases as the number of \acrshort{mpt}' parties grows.
\end{abstract}


\section{Introduction} \label{sec:introduction}
\IEEEPARstart{W}{hile} blockchains are rapidly developed and adopted in various domains, \eg, DeFi, NFT, IoT industries, contract privacy and scalability of blockchains have now become two of the top concerns. 
Unfortunately, in most existing blockchains~\cite{nakamoto2008bitcoin, wood2014ethereum}, blockchain data must be publicly accessible and verifiable so that miners can access the transaction data and re-execute transactions to verify all state transitions.

\myparagraph{Off-chain contract execution with \acrshort{mpc}} 
The demand for both privacy and scalability motivates off-chain smart contract execution frameworks. Their common idea is to offload the smart contract execution from the blockchain to off-chain systems. The blockchain then functions only as a trust anchor to verify the execution and store states. Subsequently, some promising solutions extend the off-chain contract execution to multi-party scenarios, including auction~\cite{Hawk:SP2016}, personal finance~\cite{Sinha2020LucidiTEEAT} and deal matching~\cite{GovinICBC22:Rialto, MassacciSP18:FuturesMEX}. This problem is generalized and defined in \cite{ren2021CloakDemo} as \acrfull{mpt}~\cite{ren2021CloakDemo, ren2022Cloak}. It means transitioning blockchain states by a publicly verifiable off-chain \acrshort{mpc}, where the \acrshort{mpc} takes on- and off-chain inputs from, and delivers on- and off-chain outputs to multiple parties, without leaking their inputs/outputs to the public or each other. For example, in a second-price auction~\cite{Hawk:SP2016}, multiple mutually distrustful parties jointly perform an auction on their confidential on-chain balance and off-chain bids. When the auction finishes, the party with the highest bid wins and pays the second-highest price on-chain. 
To enable \acrshort{mpt}, two kinds of solutions exist. The first is cryptography-based solutions, which adopt \acrshort{mpc}~\cite{CarstenSCN'14PAMPC, DanIACR19ZKP, cui2021mpc} or \acrfull{he}~\cite{SamuelSP2022ZeeStar} to allow parties jointly and confidentially evaluate a program off-chain, then commit the evaluation status/outputs on-chain. The second, \acrshort{tee}-based solutions~\cite{FastKitten'19, Sinha2020LucidiTEEAT, ren2021CloakDemo, ren2022Cloak}, uses \acrshort{tee} to collect private data from parties, evaluates a program with the data inside enclaves, and finally commits the evaluation status/outputs on-chain. 





\myparagraph{Limitations}
However, existing solutions of \acrshort{mpt} suffer from at least one of the following flaws: (i) Do not achieve data availability, making them vulnerable to data lost when off-chain systems fail. For example, even with \acrshort{zkp} or \acrshort{tee} to prove the correct state transitions, users cannot know their balances if an off-chain operator withholds the states. This property is keenly required by the Ethereum community~\cite{DataAvailability} and the community has designed a series of measures to uphold it, \eg, \code{calldata}~\cite{eip-4488, zkrollup, oprollup} and \code{blob}~\cite{eip-4844}, which are keys of the coming Cancun upgrade~\cite{CancunUpgrade}; (ii) Do not achieve financial fairness, so they can only assume a rate of honest nodes exists but cannot monetarily urge profit-driven nodes to behave honestly or punish the misbehaved nodes; (iii) Do not achieve delivery fairness, which requires delivering outputs to corresponding parties at almost the same time. Formally, we say a \acrshort{mpc} protocol achieves $\Delta$-fairness if the time of different parties receiving their outputs distributes in a $\Delta$-bounded period. A large $\Delta$ will lead to several attacks, \eg, a party prior to others knowing that the \acrshort{mpt} buys an ERC20 token and change the trade rate can front-run an arbitrage transaction, so-called front-running attacks, \eg, MEV~\cite{qin2022quantifying}.  (iv) Do not achieve delivery atomicity, \ie, either both committing new states and delivering outputs are guaranteed, or none of them happens. The lack of atomicity either enables adversary knowing outputs before they are being committed on-chain to abort or rewind the \acrshort{mpt}, or leads party to permanently lost their outputs when the outputs have been committed~\cite{Ekiden:2019}; (v) Require high-cost interactions with the blockchain. 




\myparagraph{Our work} In this paper, we propose \codename, a novel \acrshort{mpt}-enabled off-chain contract execution framework. \codename solves all above problems with lower gas cost and weaker assumption. Specifically, to enable \acrshort{mpt}s on a legacy blockchain, \eg, Ethereum~\cite{wood2014ethereum}, we require multiple \acrshort{tee} executors to register their \acrshort{tee}s on a deployed \codename contract. The contract thus be aware of all \acrshort{tee}s and will specify a specific \acrshort{tee} to serve all \acrshort{mpt}. Then, multiple parties can interact with the specified \acrshort{tee} off-chain to send \acrshort{mpt}s. To achieve data confidentiality and availability (\cf, i), we propose a novel data structure of commitments. The structure allows each party and \acrshort{tee} to independently access the newest states from the blockchain, even though all other entities are unavailable. To achieve financial fairness (\cf, ii) and low cost (\cf, v), we propose a novel challenge-response subprotocol. With the subprotocol, all honest entities among parties and \acrshort{tee} executors will never lose money, and at least one misbehaved entity will be punished. Especially, it enables the \codename contract to identify the misbehaviour of the specified \acrshort{tee} and replace it with another  \acrshort{tee}. To achieve atomicity (\cf, iv) and delivery fairness (\cf, iii), we require all \acrshort{tee}s to release the keys of output ciphertext only when verifying that the output commitments have been accepted and confirmed by the blockchain. This way, we achieve the complete fairness of output delivery, where multiple parties obtain their corresponding outputs in almost the same time. 
Consequently, \codename achieves the data availability, financial fairness, delivery fairness, and delivery atomicity of \acrshort{mpt}s simultaneously with only 34.4\% gas cost of the SOTA, Cloak~\cite{ren2022Cloak}, while assuming at least one parties and \acrshort{tee} executors are honest. Last, we demonstrate how to optimize or prune \codename for simpler or less secure scenarios, including how to ignore some secure properties for lower gas costs further.

\myparagraph{Contributions}
Our main contributions are as follows. 

\begin{itemize}[leftmargin=3mm, parsep=0mm, topsep=1mm, partopsep=0mm]
    \item We design a novel off-chain contract execution framework, \codename, which enables \acrshort{mpt}s on legacy blockchains.
    \item We propose a \codename protocol which handles the problem of how to maximize the security of \acrshort{mpt}s by using a minimally trusted \acrshort{tee} network. Specifically, the protocol achieves confidentiality, data availability, financial fairness, delivery fairness, and delivery atomicity simultaneously, while requiring at least one party and at least one of \acrshort{tee} executors to be honest. 
    \item We implement and evaluate \codename on 10 \acrshort{mpt}s with varying parties from 2 to 11. 
    \item We demonstrate how to optimize further or fine-tune \codename protocol to make trade-offs between security and cost for simpler scenarios.
\end{itemize}

\myparagraph{\emph{Organization}}
We organize the paper as follows. Section~\ref{sec:related-work} introduces \acrshort{mpt} and how \codename advances related work. Section~\ref{sec:overview} sketches \codename. Section~\ref{sec:protocol} details the \codename protocol. Section~\ref{sec:facilities} illustrates the implementation of \codename prototype. In Section~\ref{sec:security}, we conduct a security analysis of \codename. In Section~\ref{sec:discusion}, we discuss how to optimize the \codename protocol and make trade-offs between the security and gas cost when degenerating \acrshort{mpt} to simpler scenarios. Finally, we evaluate \codename in Section~\ref{sec:evaluation} and conclude in Section~\ref{sec:conclusion}.

\begin{table*}[!htbp]
  \centering
  \footnotesize
  \caption{\textbf{Comparing \codename with related work}. \footnotesize{The symbols \ding{53}, \ding{109}, \ding{119} and \ding{108} refer to ``non-related'', ``not-matched'', ``partially-matched'' and ``fully-matched'' respectively. ``\textbf{Adversary Model}'' means how many Byzantine entities can be tolerant. ``\textbf{Data availability}" means whether parties or \acrshort{tee}s hold \acrshort{mpt}-specific data. ``\textbf{Financial fairness}'' means honest parties never lost money while at least one misbehaved node must be punished. ``\textbf{Delivery fairness}" means either the \acrshort{mpt} fails or parties obtain their outputs in almost the same time. ``\textbf{Delivery atomicity}" means whether both committing of outputs and the delivery of output or none of them are guaranteed.}}
  \vspace{0.1cm}
  \label{tab:comparison}
  \setlength{\tabcolsep}{1.1mm}{
    \begin{tabular}{lccccccccccc}
    \toprule
    \multirow{2}{*}{\textbf{Approach}} & \multicolumn{2}{c}{\textbf{Adversary Model}} && \multirow{2}{*}{\textbf{min(\#TX)}} & \multirow{2}{*}{\textbf{Confidentiality}} && \multicolumn{2}{c}{\textbf{Data availability}} & \multirow{2}{*}{\textbf{\makecell[c]{ Financial\\Fairness }}}  & \multirow{2}{*}{\textbf{\makecell[c]{ Delivery\\Fairness }}} & \multirow{2}{*}{\textbf{\makecell[c]{ Delivery\\Atomicity }}}\\
    \cmidrule{2-3} \cmidrule{8-9}  & Parties & \acrshort{tee} Executors && & && Parties & \acrshort{tee}s & & &\\
    \midrule
    Ekiden~\cite{Ekiden:2019}& $1^*$ & $m^*-1$\tnote{1} && $O(1)$ & \ding{108} && \ding{109}\tnote{2} & \ding{109} & \ding{53} & \ding{53} & \ding{108} \\
    Confide~\cite{CONFIDE:SIGMOD20}& $1^*$ & $\lfloor (m^*-1)/2 \rfloor $\tnote{3} && $O(1)$& \ding{108} && \ding{109} & \ding{108} & \ding{53} & \ding{53} & \ding{53} \\
    POSE~\cite{FrassettoNDSS22POSE} & $1^*-1$ & $m^*-1$ && $O(1)$ & \ding{108} && \ding{119} & \ding{119} & \ding{53} & \ding{53} & \ding{109} \\
    Bhavani \etal~\cite{Choudhuri'17-FairMPC}& $n^*$ & $m^*\mid_{m=n}$ && $O(1)$ & \ding{108} && \ding{53} & \ding{53} & \ding{53} & \ding{108} & \ding{53} \\
    Hawk~\cite{Hawk:SP2016}&  $n^*$ & \ding{53} && $O(n)$ & \ding{119} && \ding{109} &\ding{53} & \ding{108} & \ding{109} & \ding{109}\\
    ZEXE~\cite{ZEXE:SP20} & $n^*$ & $1^*$ && $O(1)$ & \ding{119} && \ding{109} & \ding{53} & \ding{53} & \ding{53}\tnote{5} & \ding{109} \\
    Fastkitten~\cite{FastKitten'19} & \multicolumn{2}{c}{$(n^*+1^*)-1$}&& $O(n)$ & \ding{119}\tnote{6} && \ding{109} &\ding{109} & \ding{108} & \ding{109} & \ding{109}  \\
    LucidiTEE~\cite{Sinha2020LucidiTEEAT} & $n^*$ & $m^*-1$ && $O(n)$ & \ding{108} && \ding{109} & \ding{119} & \ding{53} & \ding{119} & \ding{108} \\
    Cloak~\cite{ren2022Cloak} & \multicolumn{2}{c}{$(n^*+1^*)-1$} && $O(1)$ & \ding{108} && \ding{109} & \ding{108} & \ding{108} & \ding{109} & \ding{109} \\
    \textbf{\codename} & $n^*-1$ & $m^*-1$ && $O(1)$ & \ding{108} && \ding{108} & \ding{108} & \ding{108} & \ding{108} & \ding{108} \\
    \bottomrule
  \end{tabular}
  \begin{tablenotes}
  \item[]
  The $^*$ denotes the total number of the specific type of entities, \eg, $1^*$ denotes the unique party/executor, $n^*$ denotes all $n$ parties, and $m^*$ denotes all executors.
  \end{tablenotes}
  }
\end{table*}

\section{Background and Related Work} \label{sec:related-work}
\subsection{\acrlong{mpt}}

Informally, \acrfull{mpt} refers to a transaction which transitions states on-chain by publicly verifiable off-chain \acrshort{mpc}. The off-chain \acrshort{mpc} in an \acrshort{mpt} takes both on-/off-chain inputs and delivers both on-/off-chain outputs. Therefore, so far, \acrshort{mpt} is the most general definition of off-chain contract execution in multi-party scenarios and can be easily applied to various domains. For example, recall the second-price auction in Section~\ref{sec:introduction}. During the process, the bids should keep private to their corresponding parties, \ie, \textit{confidentiality} is held; The public (\eg, the blockchain miners) ought to verify that the output is the correct output of a claimed joint auction, \ie, the \textit{correctness} and \textit{public verifiability} hold. We demonstrate more \acrshort{mpt} scenarios in Section~\ref{sec:evaluation}.

Formally, \acrshort{mpt} is modeled as below~\cite{ren2021CloakDemo, ren2022Cloak}. 
$$
\begin{aligned}
    & \quad  c_{s_1}, \dots, c_{s_n} \overset{c_f,~c_{x_1}, \dots, c_{x_n}}{\Longrightarrow}  c_{s'_1}, \dots, c_{s'_n}, c_{r_1}, \dots, c_{r_n}, proof \\
    & \qquad \qquad \quad  \mid s_1, \dots, s_n \stackrel{f(x_1, \dots, x_n)}{\Longrightarrow} s'_1, \dots, s'_n, r_1, \dots, r_n
\end{aligned}
$$

For a blockchain and an array of parties $\textit{\textbf{P}}$ where $|\textit{\textbf{P}}| = n~(n\in \mathbf{Z^*} \wedge n>1)$, we denote a party $\textit{\textbf{P}}[i]$ as the party $P_i$. An \acrshort{mpt} takes secret transaction parameter $x_i$ and old state $s_i$ from each $P_i$, confidentially evaluates $f$ off-chain, then delivers the secret return value $r_i$ and new state $s'_i$ to $P_i$, while publishing their commitments $c_{x_i}, c_{s_i}, c_f, c_{s'_i}, c_{r_i}$ and a $proof$ on the blockchain. 
\acrshort{mpt} should satisfy the following properties.
\begin{itemize}[leftmargin=*]
    \item \textit{Correctness}: When each $P_i$ providing $x_i, s_i$ obtains $s'_i, r_i$, it must hold that 
    $$s_1, \dots, s_n \stackrel{f(x_1, \dots, x_n)}{\Longrightarrow} s'_1, \dots, s'_n, r_1, \dots, r_n$$
    \item \textit{Confidentiality}: Each $P_i$ cannot know $\{x_j, s_j, s'_j, r_j | j\neq i\}$ except those that can be derived from public info and the secrets it provides. 
    \item \textit{Public verifiability}: With $proof$, all nodes can verify that the state transition from $\textit{\textbf{c}}_s\gets[c_{s_i}|_{1..n}]$ to $\textit{\textbf{c}}_{s'}\gets[c_{s'_i}|_{1..n}]$ is correctly caused by a unknown function $f$ (committed by $c_f$) taking unknown parameter $x_i$ (committed by $c_{x_i}$) and old state $s_i$ (committed by $c_{s_i}$), and obtains unknown new state $s'_i$ (committed by $c_{s'_i}$) and return value $r_i$ (committed by $c_{r_i}$).
\end{itemize}

\subsection{Related Work}
Here we highlights the difference and novelty of \codename, as shown in Table~\ref{tab:comparison}.


\myparagraph{\acrshort{tee}-based confidential smart contract} Ekiden~\cite{Ekiden:2019, secondstate'20}, CCF~\cite{ccf2019}, Confide~\cite{CONFIDE:SIGMOD20}, and POSE~\cite{FrassettoNDSS22POSE} are designed for confidential smart contracts where transaction inputs/outputs and contract states are confidential and all transactions are regarded as independent. These frameworks never consider properties specific in multi-party scenarios, \eg, fairness. 

Ekiden is a confidential smart contract framework which features appointing the consensus, execution, and key management functionality to different nodes. Specifically, besides consensus nodes, Ekiden sets up multiple \acrshort{tee}-enabled executors to serve users independently, where consensus nodes can obtain outputs as long as at least one executor is honest. Yet it requires executors' \acrshort{tee}s to fetch keys from a \acrshort{tee}-based key management committee to evaluate each transaction. This requirement additionally assumes the number of available \acrshort{tee} executors in the committee should be over a specific threshold, where the threshold depends on the threshold of the distributed key generation algorithms adopted by the committee. On atomicity, Ekiden proposes a \textit{two-phase protocol} which delivers keys encrypting the outputs to users off-chain when the outputs have been committed on-chain, thus achieving atomicity. On data availability, users cannot access their states on-chain since the states are encrypted by \acrshort{tee}s. However, each executor also cannot decrypt on-chain states without requesting keys from the committee. Therefore, it is flawed in data availability. 

Confide and CCF are permissioned network where \acrshort{tee}-enabled executors maintain a consensus, \eg, RAFT, thereby can tolerate less than $1/2$ unavailable executors. They store contract data (\eg, code, states) by encrypting data with keys shared among \acrshort{tee}s, thereby achieving data availability of \acrshort{tee}s. However, like Ekiden, if \acrshort{tee} executors are unavailable, users will temporarily lose accessibility to and even permanently lose their private data and on-chain assets. The \acrshort{tee}s' data availability holds.

POSE propose an off-chain contract execution which features high availability and no interaction with blockchain in optimistic cases. It introduces a challenge-response mechanism, ensuring the system's availability even if all-but-one executors are Byzantine. The protocol additionally requires the transaction sender to be honest to initiate the challenge. Users of POSE cannot access their states independently. Each \acrshort{tee} need to synchronize with other \acrshort{tee}s to obtain state updates. On atomicity, while involving reading inputs from and writing blockchain, POSE does not consider the atomicity of the on-chain writing and off-chain output delivery. 


\myparagraph{\acrshort{tee}-based smart contracts enabling \acrshort{mpt}s} 
Choudhuri \etal~\cite{Choudhuri'17-FairMPC} is the first to achieve complete fairness for general-purpose functions with the help of blockchain and \acrshort{tee}s. It requires each party to hold a \acrshort{tee} itself. Bhavani \etal does not consider executing contract relying on on-chain states, committing states on blockchains, or punishing misbehaved nodes, thus being non-related to delivery atomicity, data availability, or financial fairness.

Fastkitten~\cite{FastKitten'19} seeks to enable arbitrary contracts, especially including multi-round \acrshort{mpc}, on Bitcoin. It lets parties execute a transaction with private inputs in \acrshort{tee}s, persists the outputs locally, and only submit new state commitments with \acrshort{tee} signatures on-chain. Therefore, the party must persist all its latest private contract states and corresponding keys to ensure its ability to transition the states next time, lacking data availability. In long-running systems, parties' persisted data keep growing, making it a disaster for parties to maintain. Moreover, it involves a challenge-response mechanism to achieve financial fairness but requires each party to send a deposit transaction before each \acrshort{mpt}, leading to $O(n)$ transactions. 

LucidiTEE~\cite{Sinha2020LucidiTEEAT} loosely requires part of parties to hold \acrshort{tee}s to achieve delivery fairness. However, the time of parties receiving outputs distributes in a period the length of which equals the generation time of \acrfull{pop}\footnote{Recall that \acrshort{pop} is a proof constructed for proving that a transaction has been confirmed on a blockchain}~\cite{Cavallaro2019Tesseract, Ekiden:2019, FastKitten'19} for proving \acrshort{tee} that a key-releasing transaction has been finalized on-chain, which costs more than 50 block intervals on Ethereum~\footnote{For achieving $\leq0.001$ false negative and false positive under an adversary with $\leq1/3$ computing power of Ethereum}. Moreover, LucidiTEE requires each party to send a transaction to join an \acrshort{mpt} or deposit, leading to $O(n)$ transactions. On financial fairness, LucidiTEE lacks mechanisms to punish misbehaviours. With a similar state confidentiality mechanism as Ekiden and commitment as Fastkitten, LucidiTEE also lacks data availability.

Cloak~\cite{ren2022Cloak} firstly propose a \textit{one-deposit-multi-transaction} mechanism, where each honest party deposits coins once globally and then joins \acrshort{mpt}s ultimately. The mechanism reduces it required on-chain transactions to $O(1)$. Cloak only commits the hash of transaction data on-chain, \eg, inputs, outputs, keys, and policies. Thereby their parties also cannot access their states without \acrshort{tee} executors, \ie, lacking data availability.

\codename propose a novel commitment structure to confidentially persist states on the blockchain with low cost. Each party can access their \acrshort{mpt}-specific data from the blockchain with only its own account private key. Each \acrshort{tee} can read \acrshort{mpt}-specific data from the blockchain without the help of either parties or other \acrshort{tee}s. Consequently, even if the whole off-chain system is unavailable, the data availability of the newest states is still guaranteed. As \codename adopt the same \textit{one-deposit-multi-transaction} and a novel challenge-response protocol, it only requires $O(1)$ transactions in optimistic cases. Finally, while achieving complete delivery fairness, \codename frees parties from maintaining \acrshort{tee}s

\myparagraph{Cryptography-based smart contracts enabling \acrshort{mpt}s} 
Cryptography-based schemes usually combine \acrshort{mpc}/\acrshort{he} with \acrshort{zkp} to enable \acrshort{mpt}s. Before the combination, \acrshort{mpc}/\acrshort{he}-based works like~\cite{MPCpenalties'16, MPCBitcoin'16, PokerBitcoin'15} achieve great confidentiality but not targets public verifiability. \acrshort{zkp}-based solutions achieve public verifiability but lack confidentiality. For example, Hawk~\cite{Hawk:SP2016} requires a tight-lipped manager to collect parties' secrets, execute a contract, and generate the \acrshort{zkp} proof, thus the confidentiality of Hawk is limited. ZEXE~\cite{ZEXE:SP20} proves the satisfaction of predicates by \acrshort{zkp} proof without revealing party secrets to the public. However, generating the proof requires a party to know all predicate's secrets, thereby violating inter-party confidentiality. Combining \acrshort{mpc} with \acrshort{zkp}, public auditable \acrshort{mpc} (PA-\acrshort{mpc})~\cite{CarstenSCN'14PAMPC} achieves the publicly verifiable \acrshort{mpc}, allowing multiple parties jointly evaluate a program and prove it. Nevertheless, existing PA-\acrshort{mpc} primitives are not designed for committing data or proving state transitions, \eg, \acrshort{mpc}s expressed in Solidity that operate both on- and off-chain inputs/outputs. Moreover, they have flaws at inefficiency and weaker adversary model, and still fail in practically supporting nondeterministic negotiation or achieving financial fairness. Specifically,~\cite{CarstenSCN'14PAMPC} requires trusted setup or un-corrupted parties. \cite{FoteiniASIACRYPT20} is function-limited. \cite{OzdemirSecurity22} very recently achieves general-purpose PA-\acrshort{mpc} but only support circuit-compatible operations. None of the above solutions are for confidential smart contracts or can punish adversaries.
Instead, using the same proof structure with Cloak, \codename conforms to both confidentiality and public verifiability.
For security, while the underlining \acrshort{mpc} of~\cite{MPCpenalties'16, MPCBitcoin'16, PokerBitcoin'15, Choudhuri'17-FairMPC} requires honest-majority parties, \codename secure the system under an Byzantine adversary corrupting all parties and all-but-one \acrshort{tee} executors.

\section{\codename Design} \label{sec:overview}
In this section, we first overview the system model, adversary model, and system goals of \codename. Then, we overview \codename protocol and highlight the challenges we handled and corresponding countermeasures.

\subsection{\textbf{System model}}




Figure~\ref{fig:framework} shows the framework of \codename, \ie, a \acrshort{tee}-Blockchain system consisting of three components.

\myparagraph{Blockchain ($BC$)} A blockchain, \eg, Ethereum~\cite{wood2014ethereum}, that can deploy and evaluate Turing-complete smart contracts. 


\myparagraph{Parties ($\textit{\textbf{P}}$)} an array of parties who participate a specific \acrshort{mpt}.

\myparagraph{\codename network ($DN$)} A \codename Network consists of multiple \acrshort{tee} executors and \acrshort{tee}s, where each executor $E$ is a server hosting a \acrshort{tee} $\mathcal{E}$. We denote the set of all executors as $\textit{\textbf{E}}$ and all \acrshort{tee}s as $\mathbfcal{E}$. 






\subsection{\textbf{Adversary model}}
We assume that a Byzantine adversary presents in a \codename system. The assumptions and threats are as follows.

\myparagraph{Blockchain} We assume that \textit{BC} satisfies the common prefix, chain quality and chain growth, so it can continuously handle and reach consistency on new transactions. Moreover, there is a \acrfull{pop} scheme to prove to \acrshort{tee}s that a transaction has been finalized on \textit{BC}, which is for against eclipse attack and also adopted by~\cite{ren2022Cloak, Cavallaro2019Tesseract, FastKitten'19, Ekiden:2019}. The \acrshort{pop} of a transaction is a block sequence that contains the transaction and is provided to \acrshort{tee}s in the expected time.

\myparagraph{Parities} An honest party can access the latest view of the blockchain and trust the data it reads from the blockchain. It trusts its platform and running code but not others. An honest party also trusts the integrity and confidentiality of all \acrshort{tee}s it attested. An honest party never reveal its secrets to others except attested \acrshort{tee}s.

\myparagraph{\codename network} An honest \acrshort{tee} executor can access the latest blockchain view and trust the data it reads from the blockchain. An honest executor also trusts its platform and running code but not others. An honest executor also trusts the integrity and confidentiality of attested \acrshort{tee}s. 

\myparagraph{Threat model} A Byzantine adversary can corrupt all parties and all-but-one \acrshort{tee} executors. A corrupted party or executor can behave arbitrarily, \eg, mutating, delaying and dropping messages, but never break the integrity/confidentiality of \acrshort{tee}. Moreover, the adversary cannot interfere with the communications among honest entities, \eg, the communications among honest parties or between honest parties and honest executors.

\subsection{\textbf{System goals}}
Informally, we seek to achieve following properties. 

\myparagraph{Correctness} If an \acrshort{mpt} succeeds, the outputs must be the correct results of the \acrshort{mpt} applied to the inputs committed.

\myparagraph{Confidentiality} The inputs and outputs of \acrshort{mpt} are always confidential to their corresponding parties.

\myparagraph{Public verifiability} The public, including the blockchain, can verify the correctness of the state transition caused by a \acrshort{mpt}. Particularly, to accept a state transition, the blockchain will verify that the old states from which the new state is transitioning match its current states.

\myparagraph{Data availability} If an \acrshort{mpt} successfully completes, it holds that (i) each honest party can access the plaintext of its newest states independently, and (ii) each honest executor's \acrshort{tee} can access the plaintext of the newest states independently to restore the newest states. This means honest parties will never lose their newest states, no matter how \acrshort{tee} executors behave.

\myparagraph{Financial fairness} If at least one party is honest, then either (i) the protocol correctly completes the \acrshort{mpt} or (ii) all honest parties know that negotiation of the \acrshort{mpt} failed and stay financially neutral or (iii) all honest parties know the protocol aborted, stay financially neutral, and at least one of malicious entities must have been financially punished.

\myparagraph{Delivery fairness} If at least one \acrshort{tee} executor is honest, then either (i) all parties know the plaintext return values and new states in a $\Delta$-bounded period, or (ii) the new states and return values are not committed on-chain, and none of the parties or executors can know the plaintext of new states and return values.

\myparagraph{Delivery atomicity} If at least one \acrshort{tee} executor is honest, then either (i) some parties know the plaintext new states or return values, and the new states must have been committed on-chain, or (ii) new states are not committed on-chain, and none of the parties obtains its plaintext new states or return values.

\subsection{\textbf{Protocol workflow}}

\begin{figure}[!htb]
  \centering
  \includegraphics[width=8.3cm]{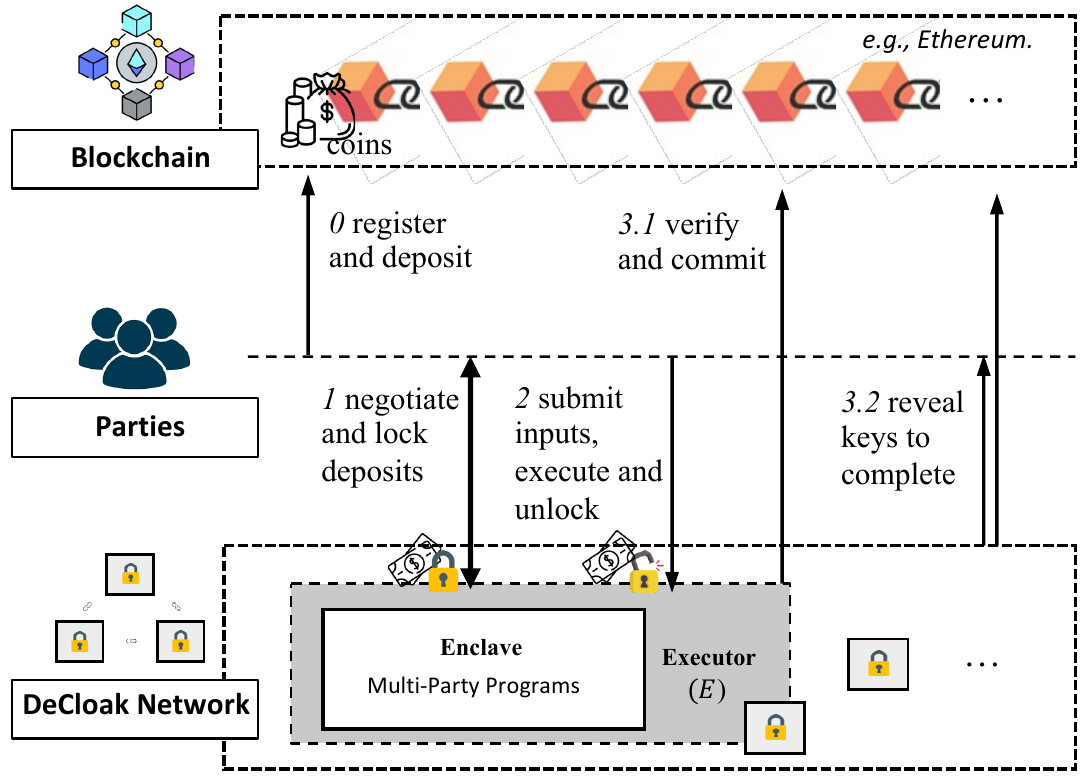}
  \caption{\textbf{The framework and workflow of \codename}.}
  \label{fig:framework}
\end{figure}

Figure~\ref{fig:framework} shows the workflow of $\pi_{\codename}$. 
We assume all \acrshort{tee}s have been registered on-chain as a \acrshort{tee} list $\mathbfcal{E}$ before the protocol started. Then, $\pi_{\codename}$ starts to serve an \acrshort{mpt} in four phases, \ie, \textit{global setup}, \textit{negotiation}, \textit{execution}, and \textit{delivery} phases. The \textit{global setup} phase happens only once for any party. Other three phases of $\pi_{\codename}$ happen in evaluating each \acrshort{mpt}.

\begin{itemize}[leftmargin=3mm, parsep=0mm, topsep=1mm, partopsep=0mm]

    \item \textbf{\textit{(0)} Global setup phase}: All parties and \acrshort{tee}s deposit some coins to the network account $ad_{\mathbfcal{E}}$ on \textit{BC}. 
    
    \item \textbf{\textit{(1)} Negotiation phase}: A party sends an \acrshort{mpt} proposal $p$ to the first executor $\mathcal{E}^*$ in the registered \acrshort{tee} executor list to initiate an \acrshort{mpt}. Upon receiving the proposal, the \acrshort{tee} $\mathcal{E}^*$ starts a \textit{nondeterministic negotiation subprotocol} $Proc_\text{nneg}$. Specifically, the $\mathcal{E}^*$ signs and broadcasts the proposal to all parties. If any party want to join or is required by the proposal, it responds with an acknowledgement to $\mathcal{E}^*$. The $\mathcal{E}^*$ keeps collecting parties' acknowledgements. When the collected acknowledgements match the settlement condition of the negotiation phase (\eg, The number of parties exceeds the number specified in the policy), $\mathcal{E}^*$ settles the proposal, deducts parties' collaterals from their coins cached in $\mathcal{E}^*$, and broadcasts the settled \acrshort{mpt} proposal $p'$ to all parties.
    
    \item \textbf{\textit{(2)} Execution phase}: Upon receiving $p'$, each party involved in the proposal submits its signed plaintext inputs (\ie, parameters) to $\mathcal{E}^*$. $\mathcal{E}^*$ first read old states on the blockchain with their \acrshort{pop}\footnote{We use the same \acrshort{pop} as~\cite{Cavallaro2019Tesseract, FastKitten'19, ren2022Cloak}}. Then, $\mathcal{E}^*$ evaluates the \acrshort{mpt} to obtain the outputs (\ie, return values and new states) inside.
    
    \item \textbf{\textit{(3.1-3.2)} Delivery phase}: When the $\mathcal{E}^*$ gets the \acrshort{mpt} outputs, it starts a \textit{$\Delta$-fair delivery subprotocol} $Proc_\text{fdel}$. First, it generates one-time symmetric keys to compute the commitments of the outputs and sends a \textit{Commit} $TX_{cmt}$ to publish output commitments on \textit{BC} with the ciphertext of the symmetric keys (encrypted by the network key $k_{\mathbfcal{E}}$). Upon $TX_{cmt}$ being confirmed on the blockchain, each $\mathcal{E}\in \mathbfcal{E}$ independently verifies the \acrshort{pop} of $TX_{cmt}$, obtains the symmetric keys from $TX_{cmt}$, then sends a $TX_{com}$ to reveal the committed outputs to each party respectively. Consequently, both parties and executors do not need to persist any \acrshort{mpt}-specific commitments or keys.
\end{itemize}

If any misbehaviour appears during the negotiation, execution, or delivery phase, we adopt a novel challenge-response mechanism to identify the misbehaved entities in parties and \acrshort{tee} executors.

\subsection{\textbf{Design challenges and highlights}}

\subsubsection{\textbf{Achieve data availability of both \acrshort{tee}s and parties}}


The challenge here is (i) how to achieve the data availability of both parties and \acrshort{tee}s and (ii) ensuring confidentiality and living in harmony with the protocol for delivery atomicity and fairness. To achieve these, we introduce a novel data commitment subprotocol. 
Specifically, say each party $P_i$ has its account $(sk_i, pk_i, ad_i)$, where $sk_i, pk_i, ad_i$ refer to the private key, public key and address of the account. As each party $P_i$ is identified by its address, we refer $P_i$ to $ad_i$ indiscriminately. We assume a common \acrshort{tee} network account $(sk_{\mathbfcal{E}}, pk_{\mathbfcal{E}}, ad_{\mathbfcal{E}})$ has been synchronized among all \acrshort{tee}s. Then, we require all entities commit private data $d_i$ on blockchain in the following structure $c_{d_i}$. $k_{d_i}$ denotes a one-time symmetric key for encrypting $d_i$. $k_{ie}$ denotes the symmetric key generated by ECDH, \ie, $k_{ie} \gets \FuncSty{ECDH}(sk_i, pk_{\mathbfcal{E}})$ and $k_{ie} \gets \FuncSty{ECDH}(sk_{\mathbfcal{E}}, pk_i)$. Consequently, on the one hand, either $P_i$ or \acrshort{tee}s can independently obtain $k_{ie}$ without interacting with the other. And a party needs only to hold the account private key $sk_i$ to access and operate its all commitments on-chain. On the other hand, when \codename release $c^*_{d_i}$ ($c_{d_i}$ without  $\FuncSty{Enc}_{k_{ie}}(k_{d_i})$) to commit and verify the state transition on-chain first for atomicity and fairness, any adversary cannot obtain $k_{d_i}$ to decrypt $\FuncSty{Enc}_{k_{d_i}}(d_i)$.
$$
c_{d_i} := [\FuncSty{Enc}_{k_{d_i}}(d_i), \FuncSty{Enc}_{k_{ie}}(k_{d_i}), P_i]
$$




\subsubsection{\textbf{Achieve complete delivery fairness}}
\label{sec:fairness-challenge}
In \codename, when a \acrshort{tee} executor evaluated the \acrshort{mpt} inside its \acrshort{tee}, the \acrshort{tee} cannot release the output immediately. Instead, the \acrshort{tee} first generates one-time symmetric keys to encrypt the outputs, then sends a $TX_{cmt}$ to publish the output ciphertext and the ciphertext of the keys on-chain. The keys' ciphertext can be decrypted by all \acrshort{tee}s independently but each \acrshort{tee} only releases the decrypted keys after $TX_{cmt}$ has been finalized on-chain. Since we assume the blockchain is ideally available, all honest \acrshort{tee} executors can feed the \acrshort{pop} of $TX_{cmt}$ to their \acrshort{tee}. Therefore, if at least one honest executor exists, parties communicating with all executors can obtain the keys to decrypt the output ciphertext at almost the same time.

\subsubsection{\textbf{Resist Byzantine adversary with minimal transactions}}

In this paper, we propose a novel \textit{challenge response subprotocol} $Proc_\text{rcha}$. At a high level, $Proc_\text{rcha}$ is designed with the following idea: when an honest party does not receive protocol messages off-chain from the specified \acrshort{tee}, it can publicly challenge the \acrshort{tee} with the proposal on-chain. The \acrshort{tee} can only avoid being punished if it can respond with expected outputs or prove that the problem is caused by some misbehaved parties rather than itself. Specifically, an \acrshort{mpt} proposal only has three possible results: (i) \code{NEGOFAILED}, which means the negotiation of the proposal failed; (ii) \code{COMPLETED}, which means the completion of the \acrshort{mpt}, \ie, an $TX_{com}$ is sent and accepted by the blockchain (iii) \code{ABORTED}, \ie, some entities misbehaved, making the \acrshort{mpt} aborted. Therefore, the challenged \acrshort{tee} needs to respond with one of the following three results to prove its honesty: (i) sending a transaction $TX_{fneg}$ to prove that the negotiation of the \acrshort{mpt} failed; (ii) sending a transaction $TX_{com}$ to complete the \acrshort{mpt} and release its outputs; (iii) sending a transaction $TX_{pnsP}$ to prove that it cannot complete the \acrshort{mpt} as expected because some parties misbehaved after the negotiation succeeded rather than itself. If none of the above transactions can be sent, the \acrshort{tee} will be punished. However, while (ii) is inherited by the success of \acrshort{mpt}, how to achieve (i) and (iii) becomes challenging. To achieve (i), we require each \acrshort{mpt} proposal should specify a block height $h_{neg}$ to notify when the negotiation phase is expected to finish. Then, a \acrshort{tee} can send a $TX_{fneg}$ to fail the proposal on-chain if it verifies that the collected acknowledgements from both off-chain $\textit{\textbf{ack}}$ and on-chain $\textit{\textbf{TX}}_{ack}$ before $h_{neg}$-th block still cannot satisfy the settlement condition of the proposal. To achieve (iii), when a \acrshort{tee} cannot complete the \acrshort{mpt}, the \acrshort{tee} needs to challenge those misbehaved parties to prove that the reason is some parties did not submit their inputs rather than itself.

\begin{table*}[!tbp]
  \centering
  \caption{A summary of main symbols}
  \footnotesize
  \label{tab:symbol}
  {
      \begin{tabular}{cccl}
        \toprule
        \textbf{Topic} & \textbf{Symbol} & \textbf{Name} & \textbf{Description} \\
        \midrule
        \multirow{4}{*}{\textbf{Framework}} 
        & $BC$ & Blockchain & \makecell[l]{A $BC$ enables Turing-complete smart contracts} \\
        & $\textbf{\textit{P}}$ & Parties & An array of an \acrshort{mpt}'s participants\\
        & $DN~(\textit{\textbf{E}}, \mathbfcal{E})$ & \codename network & A network $DN$ consisting of an array of executors $\textit{\textbf{E}}$ and \acrshort{tee}s $\mathbfcal{E}$ \\
        & $E^*$ & \acrshort{tee} executor & The server hosting the specified \acrshort{tee} $\mathcal{E}^*$ \\
        & $\mathcal{E}^*$ & \acrshort{tee} & The specified \acrshort{tee} running the enclave program $\mathcal{E}$. \\
        \cmidrule{2-4} \multirow{5}{*}{\textbf{Protocol}} 
        & $ad_{\mathbfcal{E}}, sk_{\mathbfcal{E}}$ & Enclave account & The address and private key of the common network account controlled by $\mathbfcal{E}$\\
        & $Proc_\text{nneg}$ & - & Nondeterministic negotiation subprotocol \\
        & $Proc_\text{rcha}$ & - & challenge-response subprotocol \\
        & $Proc_\text{fdel}$ & - & $\Delta$-fair delivery subprotocol \\
        \cmidrule{2-4} \multirow{9}{*}{\textbf{\acrshort{mpt}}} 
        & $TX_{chaT}$ & \textit{challengeTEE} &  A transaction from the specified \acrshort{tee} $\mathbfcal{E}[0]$ to publicly challenge the malicious parties \\
        & $TX_{ack_i}$ & \textit{acknowledge} &  A transaction from the party $P_i$ to publicly join the \acrshort{mpt} proposal \\
        & $TX_{fneg}$ & \textit{failNegotiation} &  A public response from the specified \acrshort{tee} $\mathbfcal{E}[0]$ to $TX_{chaT}$ to signal the negotiation failure \\
        & $TX_{chaP}$ & \textit{challengeParties} &  A transaction from the specified \acrshort{tee} $\mathbfcal{E}[0]$ to publicly challenge the malicious parties \\
        & $TX_{resP_i}$ & \textit{partyResponse} &  A public response from the party $P_i$ to $TX_{chaP}$\\
        & $TX_{cmt}$ & \textit{commit} &  A transaction from the specified \acrshort{tee} $\mathbfcal{E}[0]$ to commit and lock the \acrshort{mpt} outputs\\
        & $TX_{com}$ & \textit{complete}  &  A public response from the specified \acrshort{tee} $\mathbfcal{E}[0]$ to $TX_{chaT}$ to complete the \acrshort{mpt}\\
        & $TX_{pnsP}$ & \textit{punishParties} &  A public response from the specified \acrshort{tee} $\mathbfcal{E}[0]$ to $TX_{chaT}$ to punish malicious parties\\
        & $TX_{pnsT}$ & \textit{punishTEE}x &  A transaction from anyone to punish the misbehaved \acrshort{tee}\\
        \bottomrule
      \end{tabular}
  }
\end{table*}
    
\section{\codename Protocol} \label{sec:protocol}
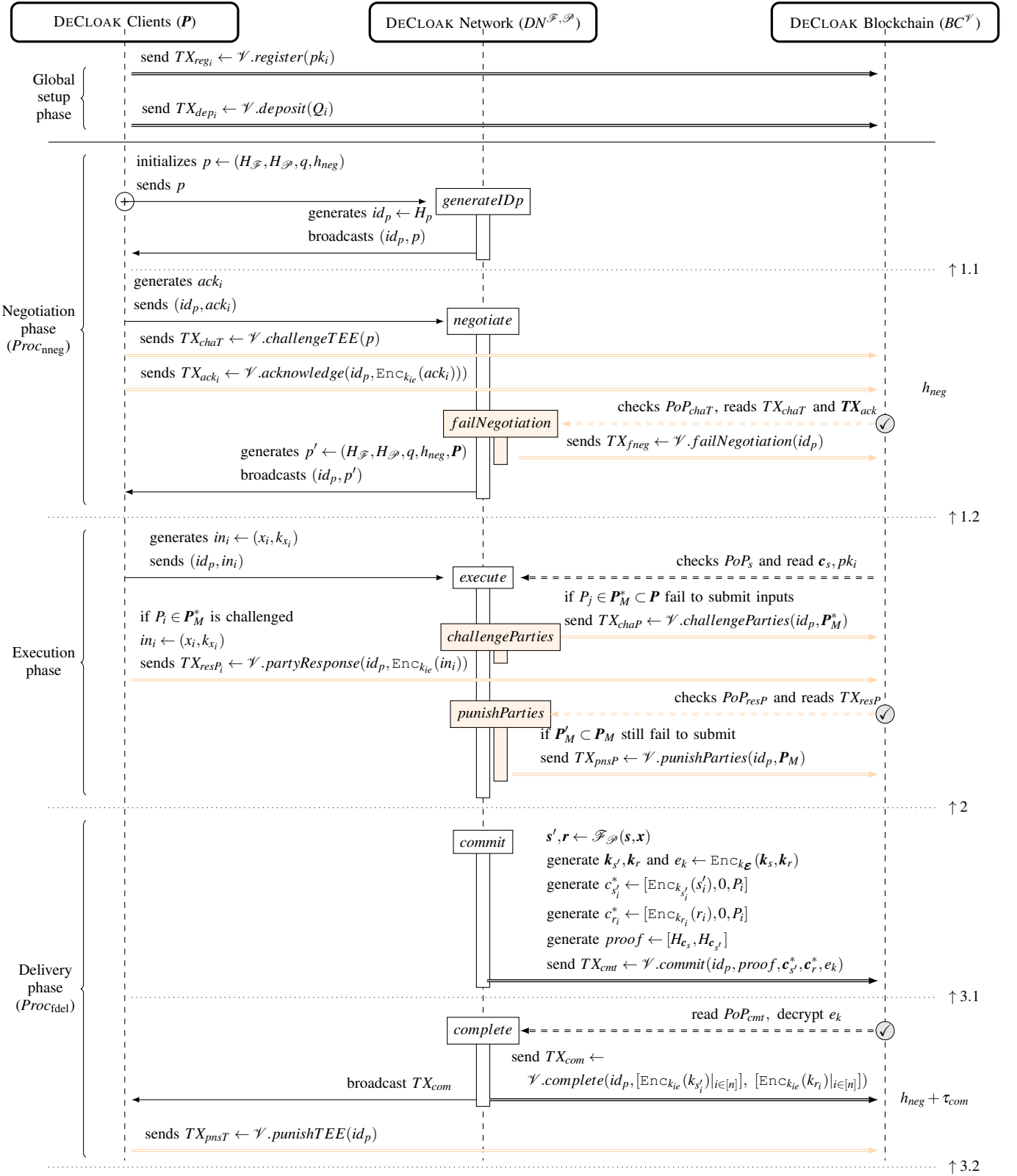
\begin{figure*}[!tbp]
    \centering
    \usetikzlibrary{calc,trees,positioning,arrows,chains,shapes.geometric,%
    decorations.pathreplacing,decorations.pathmorphing,shapes,%
    matrix,shapes.symbols, arrows.meta}
  
    \def\Initiator{Initiator}
    \def\Clients{Clients}
    \def\Executor{Network}
    \def\Blockchain{Blockchain}
    
    \def\Dividends{Dividends}
    \def\Acks{Acks}
    \def\ConfirmTx{TXs}
    \def\GetConfirmTx{GetTXs}
    \def\Proof{Proof}
    \def\PoP{PoP}
    \def\PunishExecutor{PunishExecutor}
    \def\ReleaseKey{ReleaseKey}
	
    \tikzstyle{inout}=[trapezium, trapezium left angle=60, trapezium right angle=120, draw] 
    \tikzstyle{end}=[rectangle, rounded corners, draw]   
    \tikzstyle{endn}=[rounded rectangle, draw]   
    \tikzstyle{exec}=[rectangle, draw]    
    \tikzstyle{decide}=[kite, kite vertex angles=120, draw]   

    \tikzset{
    >=stealth',
      punktchain/.style={
        rectangle, 
        rounded corners, 
        draw=black, very thick,
        text width=10em, 
        minimum height=3em, 
        text centered, 
        on chain},
      line/.style={draw, thick, <-},
      element/.style={
        tape,
        top color=white,
        bottom color=blue!50!black!60!,
        minimum width=8em,
        draw=blue!40!black!90, very thick,
        text width=10em, 
        minimum height=3.5em, 
        text centered, 
        on chain},
      every join/.style={->, thick,shorten >=1pt},
      decoration={brace},
      tuborg/.style={decorate},
      tubnode/.style={midway, left=2pt},
        >=stealth',
        punkt/.style={
               rectangle,
               rounded corners,
               draw=black, very thick,
               text width=11em,
               minimum height=2em,
               text centered},
        pil/.style={
           ->,
           thick,
           shorten <=2pt,
           shorten >=2pt,}
    }
	
    \begin{tikzpicture}[every node/.style={font=\footnotesize,
      minimum height=0.1cm,minimum width=0.1cm}]
        \node [matrix, very thin, column sep=1.9cm, row sep=0.7cm] (matrix) at (0,0) {
            &[-4ex] \node(0,0) (\Clients) {}; &[14ex] & \node(0,0) (\Executor) {}; &[19ex] & \node(0,0) (\Blockchain) {}; &[-7ex]  \\
            [0ex] & \node(0,0) (\Clients G0) {}; & \node(0,0) (\ConfirmTx G0p-b) {};  & \node(0,0) (\Executor G0) {}; & & \node(0,0) (\Blockchain G0) {};  & \\
            [0ex]& \node(0,0) (\Clients G1) {}; & \node(0,0) (\ConfirmTx G1p-b) {};  & \node(0,0) (\Executor G1) {}; & & \node(0,0) (\Blockchain G1) {};  & \\
            [-4ex] \node(0,0) (tg1 left) {}; & \node(0,0) (\Clients tg1-1) {};  & & \node(0,0) (\Executor tg1-1) {};  & & & \node(0,0) (tg1 right) {}; \\
            [-2ex] & \node(0,0) (\Clients 0) {}; &  & \node(0,0) (\Executor 0) {}; & & \node(0,0) (\Blockchain 0) {};  & \\
            [-3ex] \node(0,0) (t0 left) {}; & \node(0,0) (\Clients 0-1) {};  & \node(0,0) (\Dividends 0-1) {}; & \node(0,0) (\Executor 0-1) {};  & & & \node(0,0) (t0 right) {}; \\
            & \node(0,0) (\Clients 1) {}; & \node(0,0) (\Dividends 1) {}; & \node(0,0) (\Executor 1) {}; & & & \\
            [-4ex] \node(0,0) (t1 left) {}; &\node(0,0) (\Clients 1-1) {};  & \node(0,0) (\Dividends 1-1) {};  & \node(0,0) (\Executor 1-1) {};  & & & \node(0,0) (t1 right) {};\\
            [0ex]& \node(0,0) (\Clients 2) {}; & \node(0,0) (\Acks) {}; & \node(0,0) (\Executor 2) {}; &  \node(0,0) (\GetConfirmTx 2) {};  &  \node(0,0) (\Blockchain 2) {};  & \\
            [-2ex] \node(0,0) (t4-01 left) {}; & \node(0,0) (\Clients 4-01) {}; & \node(0,0) (\ConfirmTx challengeTEE) {};   & \node(0,0) (\Executor 4-01) {}; & & \node(0,0) (\Blockchain 4-01) {}; & \node(0,0) (t4-01 right) {};  \\
            [-2ex] \node(0,0) (t4-001 left) {}; & \node(0,0) (\Clients 4-001) {}; & \node(0,0) (\ConfirmTx acknowledge) {};   & \node(0,0) (\Executor 4-001) {}; & & \node(0,0) (\Blockchain 4-001) {}; & \node(0,0) (t4-001 right) {};  \\
            [-2ex] \node(0,0) (t3-0 left) {}; & \node(0,0) (\Clients 3-0) {}; & & \node(0,0) (\Executor 3-0) {}; & \node(0,0) (\GetConfirmTx acks) {};  & \node(0,0) (\Blockchain 3-0) {};  & \node(0,0) (t3-0 right) {};  \\
            [-2ex] & \node(0,0) (\Clients 3-00) {}; & & \node(0,0) (\Executor 3-00) {}; & \node(0,0) (\ConfirmTx failNegotiation) {};  & \node(0,0) (\Blockchain 3-00) {}; &  \\
            [-2ex] & \node(0,0) (\Clients 3) {}; & \node(0,0) (\ConfirmTx proposal) {};  & \node(0,0) (\Executor 3) {}; &  & \node(0,0) (\Blockchain 3) {}; & \\
            [-3ex] \node(0,0) (t3 left) {}; & \node(0,0) (\Clients 3-1) {};  & & & & & \node(0,0) (t3 right) {}; \\
            [1ex] & \node(0,0) (\Clients 4) {}; & \node(0,0) (\Dividends 4) {}; & \node(0,0) (\Executor 4) {}; & \node(0,0) (\GetConfirmTx 4)  {}; & \node(0,0) (\Blockchain 4) {}; &  \\
            [-4ex] \node(0,0) (t4-0 left) {}; & \node(0,0) (\Clients 4-0) {}; & \node(0,0) (\Dividends 4-0) {}; & \node(0,0) (\Executor 4-0) {}; &  & \node(0,0) (\Blockchain 4-0) {}; & \node(0,0) (t4-0 right) {}; \\
            [-1ex] \node(0,0) (t4-1 left) {}; & \node(0,0) (\Clients 4-1) {}; & & \node(0,0) (\Executor 4-1) {}; & \node(0,0) (\ConfirmTx challengeParties) {};  & \node(0,0) (\Blockchain 4-1) {}; & \node(0,0) (t4-1 right) {};  \\
            [-1ex] \node(0,0) (t4-2 left) {}; & \node(0,0) (\Clients 4-2) {}; & \node(0,0) (\ConfirmTx response) {};  & \node(0,0) (\Executor 4-2) {}; &  & \node(0,0) (\Blockchain 4-2) {};  & \node(0,0) (t4-2 right) {};  \\
            [-2ex] \node(0,0) (t4-3 left) {}; & \node(0,0) (\Clients 4-3) {}; & & \node(0,0) (\Executor 4-3) {}; & \node(0,0) (\GetConfirmTx response) {};  & \node(0,0) (\Blockchain 4-3) {};  & \node(0,0) (t4-3 right) {};  \\
            [1ex] & \node(0,0) (\Clients 5) {}; & & \node(0,0) (\Executor 5) {}; & \node(0,0) (\ConfirmTx punish) {};  & \node(0,0) (\Blockchain 5) {}; &  \\
            [-4ex] \node(0,0) (t5-2 left) {}; & \node(0,0) (\Clients 5-2) {}; & & \node(0,0) (\Executor 5-2) {};  &  & \node(0,0) (\Blockchain 5-2) {}; & \node(0,0) (t5-2 right) {}; \\
            [-4ex] \node(0,0) (p2 left) {}; & \node(0,0) (\Clients p2) {}; & & \node(0,0) (\Executor p2) {};  &  & \node(0,0) (\Blockchain p2) {}; & \node(0,0) (p2 right) {}; \\
            [-2ex] & \node(0,0) (\Clients 5-2-0) {}; & \node(0,0) (\Dividends 5-2-0) {}; & \node(0,0) (\Executor 5-2-0) {}; & \node(0,0) (\Proof 5-2-0) {};  & \node(0,0) (\Blockchain 5-2-0) {}; & \\ 
            [10ex] & \node(0,0) (\Clients 5-2-1) {}; & \node(0,0) (\Dividends 5-2-1) {}; & \node(0,0) (\Executor 5-2-1) {}; & \node(0,0) (\Proof 5-2-1) {};  & \node(0,0) (\Blockchain 5-2-1) {}; & \\ 
            [-4ex]\node(0,0) (t5-3 left) {}; & \node(0,0) (\Clients 5-3) {}; & & \node(0,0) (\Executor 5-3) {};  &  & \node(0,0) (\Blockchain 5-3) {}; & \node(0,0) (t5-3 right) {}; \\ 

            [-4ex] \node(0,0) (t7-1 left) {}; & \node(0,0) (\Clients 7-1) {};  & & \node(0,0) (\Executor 7-1) {}; &  & \node(0,0) (\Blockchain 7-1) {};  & \node(0,0) (t7-1 right) {}; \\
            [-8ex] & \node(0,0) (\Clients 8) {}; & \node(0,0) () {}; & \node(0,0) (\Executor 8) {}; & \node(0,0) (\PoP com) {};   & \node(0,0) (\Blockchain 8) {}; & \\
            [2ex] \node(0,0) (t8-1 left) {}; & \node(0,0) (\Clients 8-1) {};  & \node(0, 0) (\ReleaseKey) {}; & \node(0,0) (\Executor 8-1) {};  & \node(0, 0) (\ConfirmTx complete) {};  & \node(0,0) (\Blockchain 8-1) {};  & \node(0,0) (t8-1 right) {}; \\
            [-4ex] \node(0,0) (t8-2 left) {}; & \node(0,0) (\Clients 8-2) {};  & & \node(0,0) (\Executor 8-2) {};  & & \node(0,0) (\Blockchain 8-2) {};  & \node(0,0) (t8-2 right) {}; \\
            [-2ex]  & \node(0,0) (\Clients 7-0) {};  & \node(0,0) (\PunishExecutor) {};  & \node(0,0) (\Executor 7-0) {}; &  & \node(0,0) (\Blockchain 7-0) {};  & \\
            [-4ex] \node(0,0) (t9 left) {};  & \node(0,0) (\Clients 9) {}; & & \node(0,0) (\Executor 9) {}; & & \node(0,0) (\Blockchain 9) {}; & \node(0,0) (t9 right) {}; \\
        };
        
        \fill 
        	(\Clients) node[punkt] {\codename \Clients~($\textit{\textbf{P}}$)}
        	(\Executor) node[punkt] {\codename \Executor~($DN^{\mathcal{F}, \mathcal{P}}$)}
            (\Blockchain) node[punkt] {\codename \Blockchain~($BC^{\mathcal{V}}$)};
            
        \draw [dotted] 
          (\Clients 1-1) -- (t1 right) node[right] {$\uparrow 1.1$}
          (t3 left) -- (t3 right) node[right] {$\uparrow 1.2$}
          (p2 left) -- (p2 right) node[right] {$\uparrow 2$ }
          (\Clients 5-3) -- (t5-3 right) node[right,rotate=0] {$\uparrow 3.1$}
          (t9 left) -- (t9 right) node[right] {$\uparrow 3.2$};
          
        \draw [dashed] 
          (\Clients) -- (\Clients 9)
          (\Executor) -- (\Executor 9)
          (\Blockchain) -- (\Blockchain 9);
        
        \draw
            (tg1 left) -- (tg1 right) node[right] {};
        
        \filldraw[fill=white]
            (\Executor 0-1.north west) rectangle (\Executor 1.south east)
            (\Executor 2.north west) rectangle (\Executor 3.south east)
            (\Executor 4.north west) rectangle (\Executor 5-2.south east)
            (\Executor 5-2-0.north west) rectangle (\Executor 5-2-1.south east)
            (\Executor 8.north west) rectangle (\Executor 8-1.south east);
    
        \filldraw[fill=orange!10]
            let \p1=(\Executor 3-0.north west), \p2=(\Executor 3-00.south east) in ($(\x1+0.9em, \y1)$) rectangle ($(\x2+0.9em, \y2)$)
            (\Executor 3-0) node[draw, fill=orange!10, xshift=0.9em] {$failNegotiation$}
            let \p1=(\Executor 4-1.north west), \p2=(\Executor 4-1.south east) in ($(\x1+0.9em, \y1)$) rectangle ($(\x2+0.9em, \y2-1em)$)
            (\Executor 4-1) node[draw, fill=orange!10, xshift=0.9em] {$challengeParties$}
            let \p1=(\Executor 4-3.north west), \p2=(\Executor 5.south east) in ($(\x1+0.9em, \y1)$) rectangle ($(\x2+0.9em, \y2)$)
            (\Executor 4-3) node[draw, fill=orange!10, xshift=0.9em] {$punishParties$};
            
        \draw
            (\Clients 0-1) node[draw,circle,fill=white!20] {} node {+}
            (\Executor 0-1) node[draw, fill=white] {$generateIDp$}
            (\Executor 2) node[draw, fill=white] {$negotiate$}
            (\Executor 4) node[draw, fill=white] {$execute$}
            (\Executor 5-2-0) node[draw, fill=white] {$commit$}
            (\Executor 8) node[draw, fill=white] {$complete$};
            
        \draw [-latex] let \p1=(\Clients 0-1), \p2=(\Executor 0-1) in ($(\x1, \y1)$) -- ($(\x2-2.9em, \y2)$);
        \draw [-latex] (\Executor 1.west) -- (\Clients 1);
        \draw [-latex] let \p1=(\Clients 2), \p2=(\Executor 2) in ($(\x1, \y1)$) -- ($(\x2-2em, \y2)$);
        \draw [-latex] let \p1=(\Executor 3.west), \p2=(\Clients 3) in ($(\x1, \y1)$) -- ($(\x2, \y2)$);
        \draw [-latex] let \p1=(\Clients 4), \p2=(\Executor 4) in ($(\x1, \y1)$) -- ($(\x2-2em, \y2)$);
        \draw [-latex] (\Executor 8-1) -- (\Clients 8-1);
        
        \draw [line width=0.5pt, double distance=0.5pt, arrows = {-Latex[length=1.5pt 2 0]}] (\Clients G0) -- (\Blockchain G0);
        \draw [line width=0.5pt, double distance=0.5pt, arrows = {-Latex[length=1.5pt 2 0]}] (\Clients G1) -- (\Blockchain G1);
        \draw [line width=0.5pt, double distance=0.5pt, arrows = {-latex}, color=orange!30] let \p1=(\Clients 4-01), \p2=(\Blockchain 4-01.west) in ($(\x1, \y1)$) -- ($(\x2, \y2)$); 
        \draw [line width=0.5pt, double distance=0.5pt, arrows = {-latex}, color=orange!30] let \p1=(\Clients 4-001), \p2=(\Blockchain 4-001.west) in ($(\x1, \y1)$) -- ($(\x2, \y2)$); 
        \draw [line width=0.5pt, double distance=0.5pt, arrows = {-latex}, color=orange!30] let \p1=(\Executor 3-00.east), \p2=(\Blockchain 3-00.west) in ($(\x1+1.5em, \y1)$) -- ($(\x2, \y2)$); 
        \draw [line width=0.5pt, double distance=0.5pt, arrows = {-latex}, color=orange!30] let \p1=(\Executor 4-1), \p2=(\Blockchain 4-1.west) in ($(\x1+4em, \y1)$) -- ($(\x2, \y2)$);
        \draw [line width=0.5pt, double distance=0.5pt, arrows = {-latex}, color=orange!30] (\Clients 4-2) -- (\Blockchain 4-2);
        \draw [line width=0.5pt, double distance=0.5pt, arrows = {-latex}, color=orange!30] let \p1=(\Executor 5), \p2=(\Blockchain 5.west) in ($(\x1+1.5em, \y1)$) -- ($(\x2, \y2)$); 
        \draw [line width=0.5pt, double distance=0.5pt, arrows = {-latex}] let \p1=(\Executor 5-2-1), \p2=(\Blockchain 5-2-1.west) in ($(\x1+0.2em, \y1)$) -- ($(\x2, \y2)$); 
        \draw [line width=0.5pt, double distance=0.5pt, arrows = {-latex}, color=orange!30] (\Clients 7-0) -- (\Blockchain 7-0);
        \draw [line width=0.5pt, double distance=0.5pt, arrows = {-latex}] (\Executor 8-1) -- (\Blockchain 8-1);
        
        \draw [dashed, double distance=0.5pt, arrows = {-latex}, color=orange!30] let \p1=(\Blockchain 3-0), \p2=(\Executor 3-0) in ($(\x1, \y1)$) -- ($(\x2+4em, \y2)$); 
        \draw [dashed, double distance=0.5pt, arrows = {-latex}] let \p1=(\Blockchain 4), \p2=(\Executor 4) in ($(\x1-0.6em, \y1)$) -- ($(\x2+1.9em, \y2)$);
        \draw [dashed, double distance=0.5pt, arrows = {-latex}, color=orange!30] let \p1=(\Blockchain 4-3), \p2=(\Executor 4-3) in ($(\x1, \y1)$) -- ($(\x2+3.4em, \y2)$); 
        \draw [dashed, double distance=0.5pt, arrows = {-latex}] let \p1=(\Blockchain 8), \p2=(\Executor 8) in ($(\x1-0.6em, \y1)$) -- ($(\x2+1.9em, \y2)$);
        
        \draw[tuborg, decoration={brace}] let \p1=(\Clients tg1-1.south), \p2=(\Clients G0.north) in ($(\x1-2em, \y1+1em)$) -- ($(\x1-2em, \y2)$) node[tubnode] {\makecell[c]{Global\\setup\\phase}};
        \draw[tuborg, decoration={brace}] let \p1=(\Clients 3-1.south), \p2=(\Clients tg1-1.north) in ($(\x1-2em, \y1+1em)$) -- ($(\x1-2em, \y2-1em)$) node[tubnode] {\makecell[c]{Negotiation\\phase\\($Proc_\text{nneg}$)}};
        \draw[tuborg, decoration={brace}] let \p1=(\Clients p2.south), \p2=(\Clients 3-1.north) in ($(\x1-2em, \y1+1em)$) -- ($(\x1-2em, \y2-1em)$) node[tubnode] {\makecell[c]{Execution\\phase}};
        \draw[tuborg, decoration={brace}] let \p1=(\Clients 7-0.south), \p2=(\Clients p2.north) in ($(\x1-2em, \y1)$) -- ($(\x1-2em, \y2-1em)$) node[tubnode] {\makecell[c]{Delivery\\phase\\($Proc_\text{fdel}$)}};
        
        \draw
            (\Blockchain 3-0) node[draw,circle,fill=gray!20] {} node {\checkmark}
            (\Blockchain 4-3) node[draw,circle,fill=gray!20] {} node {\checkmark}
            (\Blockchain 8) node[draw,circle,fill=gray!20] {} node {\checkmark};

        \fill
            (\ConfirmTx G0p-b) 
                node[above, shift={(0:-2.3cm)}] {
                    \begin{math}
                        \begin{aligned}
                            \text{send}~TX_{reg_i} \gets \mathcal{V}.register(pk_i)
                        \end{aligned}
                    \end{math}
                }
                node[font=\footnotesize, below] {}
            (\ConfirmTx G1p-b) 
                node[above, shift={(0:-2.3cm)}] {
                    \begin{math}
                        \begin{aligned}
                            \text{send}~TX_{dep_i} \gets \mathcal{V}.deposit(Q_i)
                        \end{aligned}
                    \end{math}
                }
                node[font=\footnotesize, below] {}
            (\Dividends 0-1)  
                node[above, shift={(0:-2.25cm)}] {
                    \begin{math}
                        \begin{aligned}
                            & \text{initializes}~p \gets (H_{\mathcal{F}},H_{\mathcal{P}}, q, h_{neg}) \\
                            & \text{sends}~p
                        \end{aligned}
                    \end{math}
                }
                node[font=\footnotesize, below] {}
            (\Dividends 1) 
                node[above, shift={(0:0.1cm)}] {
                    \begin{math}
                        \begin{aligned}
                            & \text{generates}~id_p  \gets  H_p \\
                            & \text{broadcasts}~( id_p, p )
                        \end{aligned}
                    \end{math}
                }
                node[font=\footnotesize, below] {}
            (\Acks) 
                node[above, shift={(0:-3.25cm)}] {
                    \begin{math}
                        \begin{aligned}
                            & \text{generates}~ack_i\\
                            & \text{sends}~(id_p, ack_i)
                        \end{aligned}
                    \end{math}
                }
                node[font=\footnotesize, below] {}
            (\ConfirmTx proposal) 
                node[above, shift={(0:-0.2cm)}] {
                    \begin{math}
                        \begin{aligned}
                            & \text{generates}~p' \gets (H_{\mathcal{F}}, H_{\mathcal{P}}, q, h_{neg}, \textit{\textbf{P}}) \\
                            & \text{broadcasts}~(id_p, p') 
                        \end{aligned}
                    \end{math}
                }
                node[font=\footnotesize, below] {}
            (\ConfirmTx failNegotiation) 
                node[above, shift={(0:-1.3cm)}] {
                    \begin{math}
                        \begin{aligned}
                            \text{sends}~TX_{fneg}\gets \mathcal{V}.failNegotiation(id_p) 
                        \end{aligned}
                    \end{math}
                }
                node[font=\footnotesize, below] {}
            (\Blockchain 4-001) 
                node[shift={(0:0.9cm)}] {
                    $h_{neg}$
                }
            (\ConfirmTx challengeTEE) 
                node[above, shift={(0:-1.9cm)}] {
                    \begin{math}
                        \begin{aligned}
                            \text{sends}~TX_{chaT} \gets \mathcal{V}.challengeTEE(p) 
                        \end{aligned}
                    \end{math}
                }
                node[font=\footnotesize, below] {}
            (\ConfirmTx acknowledge) 
                node[above, shift={(0:-1.1cm)}] {
                    \begin{math}
                        \begin{aligned}
                            \text{sends}~TX_{ack_i} \gets \mathcal{V}.acknowledge(id_p, \FuncSty{Enc}_{k_{ie}}(ack_i)))
                        \end{aligned}
                    \end{math}
                }
                node[font=\footnotesize, below] {}
            (\ConfirmTx challengeParties) 
                node[above, shift={(0:-1.1cm)}] {
                    \begin{math}
                        \begin{aligned}
                            & \text{if}~P_j\in\textit{\textbf{P}}^*_M\subset \textit{\textbf{P}}~\text{fail to submit inputs} \\
                            & \text{send}~TX_{chaP} \gets \mathcal{V}.challengeParties(id_p, \textit{\textbf{P}}^*_M) 
                        \end{aligned}
                    \end{math}
                }
                node[font=\footnotesize, below] {}
            (\ConfirmTx response) 
                node[above, shift={(0:-1.1cm)}] {
                    \begin{math}
                        \begin{aligned}
                            & \text{if}~P_i\in \textit{\textbf{P}}^*_M~\text{is challenged} \\
                            & in_i \gets ( x_i, k_{x_i}) \\
                            & \text{sends}~TX_{resP_i} \gets \mathcal{V}.partyResponse(id_p, \FuncSty{Enc}_{k_{ie}}(in_i))
                        \end{aligned}
                    \end{math}
                }
                node[font=\footnotesize, below] {}
            (\GetConfirmTx acks) 
                node[above, shift={(0:-0.3cm)}] {
                    \begin{math}
                        \begin{aligned}
                            & \text{checks}~PoP_{chaT}\text{, reads}~TX_{chaT}~\text{and}~\textit{\textbf{TX}}_{ack}~\\
                        \end{aligned}
                    \end{math}
                }
                node[font=\footnotesize, below] {}   
            (\GetConfirmTx response) 
                node[above, shift={(0:0.2cm)}] {
                    \begin{math}
                        \begin{aligned}
                            & \text{checks}~PoP_{resP}~\text{and reads}~TX_{resP}\\
                        \end{aligned}
                    \end{math}
                }
                node[font=\footnotesize, below] {}    
            (\ConfirmTx punish) 
                node[above, shift={(0:-1.7cm)}] {
                    \begin{math}
                        \begin{aligned}
                            & \text{if}~\textit{\textbf{P}}'_M\subset\textit{\textbf{P}}_M~\text{still fail to submit} \\
                            & \text{send}~TX_{pnsP} \gets \mathcal{V}.punishParties(id_p, \textit{\textbf{P}}_M)\\
                        \end{aligned}
                    \end{math}
                }
                node[font=\footnotesize, below] {}
            (\Dividends 4)  
                node[above, shift={(0:-2.55cm)}] {
                    \begin{math}
                        \begin{aligned}
                            & \text{generates}~in_i \gets ( x_i, k_{x_i} ) \\
                            & \text{sends}~(id_p, in_i)\\
                        \end{aligned}
                    \end{math}
                }
                node[font=\footnotesize, below] {}
            (\GetConfirmTx 4)  
                node[above, shift={(0:0cm)}] {
                    \begin{math}
                        \begin{aligned}
                            \text{checks}~PoP_{s}~\text{and read}~\textit{\textbf{c}}_{s}, pk_i
                        \end{aligned}
                    \end{math}
                }
                node[font=\footnotesize, below] {}
            (\Proof 5-2-1)
                node[above, shift={(0cm:-1.3cm)}] {
                    \begin{math}
                        \begin{aligned}
                            & \textit{\textbf{s}}', \textit{\textbf{r}} \gets \mathcal{F}_{\mathcal{P}}(\textit{\textbf{s}}, \textit{\textbf{x}}) \\
                            & \text{generate}~\textit{\textbf{k}}_{s'}, \textit{\textbf{k}}_r~\text{and}~e_k \gets \FuncSty{Enc}_{k_{\mathbfcal{E}}}(\textit{\textbf{k}}_s, \textit{\textbf{k}}_r) \\
                            & \text{generate}~c^*_{s'_i} \gets [\FuncSty{Enc}_{k_{s'_i}}(s'_i), 0, P_i] \\
                            & \text{generate}~c^*_{r_i} \gets [\FuncSty{Enc}_{k_{r_i}}(r_i), 0, P_i] \\
                            & \text{generate}~proof \gets [H_{\textit{\textbf{c}}_{s}}, H_{\textit{\textbf{c}}_{s'}}] \\
                            &\text{send}~TX_{cmt} \gets \mathcal{V}.commit(id_p, proof, \textit{\textbf{c}}^*_{s'}, \textit{\textbf{c}}^*_r, e_k)
                        \end{aligned}
                    \end{math}
                }
                node[font=\footnotesize, below] {}
            (\PoP com)
                node[above, shift={(0:0cm)}] {
                    \begin{math}
                        \begin{aligned}
                            \text{read}~PoP_{cmt},~\text{decrypt}~e_k
                        \end{aligned}
                    \end{math}
                }
                node[font=\footnotesize, below] {}
            (\ReleaseKey) 
                node[above, shift={(0:0.8cm)}] {
                    \begin{math}
                        \begin{aligned}
                            & \text{broadcast}~TX_{com}
                        \end{aligned}
                    \end{math}
                }
                node[font=\footnotesize, below] {}
            (\ConfirmTx complete) 
                node[above, shift={(0:-1.4cm)}] {
                    \begin{math}
                        \begin{aligned}
                            & \text{send}~TX_{com} \gets \\
                            & \quad \mathcal{V}.complete(id_p, [\FuncSty{Enc}_{k_{ie}}(k_{s'_i})|_{i\in [n]}],~[\FuncSty{Enc}_{k_{ie}}(k_{r_i})|_{i\in [n]}])
                        \end{aligned}
                    \end{math}
                }
                node[font=\footnotesize, below] {}
            (\Blockchain 8-1) 
                node[shift={(0:0.9cm)}] {
                    $h_{neg} + \tau_{com}$
                }
            (\PunishExecutor) 
                node[above, shift={(0:-1.7cm)}] {
                    \begin{math}
                        \begin{aligned}
                          & \text{sends}~TX_{pnsT} \gets \mathcal{V}.punishTEE(id_p)
                        \end{aligned}
                    \end{math}
                }
                node[font=\footnotesize, below] {};
                
    \end{tikzpicture}
    
    \caption{\textbf{The \codename protocol $\pi_{\codename}$}. \small{The $DN^{\mathcal{F}, \mathcal{P}}$ denotes a \codename Network in which all executors hold \acrshort{tee}s with deployed $\mathcal{F}, \mathcal{P}$. 
    $BC^{\mathcal{V}, \mathcal{V}}$ denotes a blockchain with deployed \codename contract $\mathcal{V}$. 
    $Proc_\text{nneg}$ and $Proc_\text{fdel}$ denote the nondeterministic negotiation, and $\Delta$-fair delivery subprotocols, respectively. 
    Double dashed arrows denote reading $BC$ and double arrows denote writing $BC$. 
    Orange arrows denote the messages of challenge-response. 
    Other arrows denote off-chain communications in secure channels. Specifically, messages sent by parties are signed by parties and encrypted by $k_{ie}$ of $DN$, where $k_{ie}\gets\FuncSty{ECDH}(sk_i, pk_{\mathbfcal{E}})$. All messages broadcast by $DN$ are plaintext in default and signed by $sk_{\mathbfcal{E}}$. For simplicity, we omit marking ciphertext of messages that parties are sending to $DN$, but mark the ciphertext explicitly in each transaction sent to $BC$.}}
    \label{prot:decloak}
\end{figure*}

In this section, we present the \codename protocol $\pi_{\codename}$ in detail. Given a blockchain $BC$, a \codename Network $DN$ having an array of executors $\textit{\textbf{E}}$ and \acrshort{tee}s $\mathbfcal{E}$, we assume a common network account $(sk_{\mathbfcal{E}}, pk_{\mathbfcal{E}}, ad_{\mathbfcal{E}})$ has been synchronized among all \acrshort{tee}s $\mathbfcal{E}$. For an \acrshort{mpt} $\mathcal{F}$ with its party set $\textit{\textbf{P}}$, we assume $|\textit{\textbf{E}}|=|\mathbfcal{E}|=m$ and $|\textit{\textbf{P}}|=n$. Since $\pi_{\codename}$ involves data from different parties, we use $d_i$ 
to denote the private data of $P_i$ (\eg, $x_i, s_i, k_{s_i}$), $\textit{\textbf{d}}$ to denote an array $[d_i|_{1..n}]$ including all $d_i$ from $n$ parties (\eg, $\textit{\textbf{x}}, \textit{\textbf{s}}, \textit{\textbf{k}}_s$). We let $H_{d_i}$ denote $\FuncSty{hash}(d_i)$ and $H_{d}$ denote $[\FuncSty{hash}(d_i)|_{1..n}]$ (\eg, $H_{\textit{\textbf{c}}_{x}}$ denotes the hash of the array of transaction parameters $[\FuncSty{hash}(c_{x_i})|_{1..n}]$). The main symbols we will use are summarized in Table~\ref{tab:symbol}. Next, we picture the whole protocol in Figure~\ref{prot:decloak}.



\subsection{Global setup phase}
Before evaluating any \acrshort{mpt}, each party $P_i$ is supposed to $register$ their account public key $pk_i$ and $deposit$ some coins with amount $Q_i$ to the \codename contract $\mathcal{V}$ (Algorithm~\ref{alg:cloak-service}). We stress that each party only needs to do it once.

\subsection{Negotiation phase} An \acrshort{mpt} is started from its negotiation phase, where \codename uses the \textit{nondeterministic negotiation subprotocol} ($Proc_\text{nneg}$) to guide parties to reach a agreement on an \acrshort{mpt} proposal. In detail, $Proc_\text{nneg}$ proceeds in two steps. 

\textit{\textbf{1.1}}: A party who wants to call an \acrshort{mpt} $\mathcal{F}$ sends an \acrshort{mpt} proposal $p = (\mathcal{F}, \mathcal{P}, q, h_{neg})$ to the first executor $E^*$ in the registered \acrshort{tee} executor list, \ie, $E^*= \textit{\textbf{E}}[0]$, to initiate an \acrshort{mpt}. Sending proposals to other \acrshort{tee}s will be rejected by the \acrshort{tee}s. $\mathcal{P}$ denotes a privacy policy of $\mathcal{F}$. Briefly, $\mathcal{P}$ captures what data are needed by the \acrshort{mpt} $\mathcal{F}$ and how to confide these data. We detail and formalize the $\mathcal{P}$ in Appendix~\ref{sec:model}. $q$ denotes the collateral required for joining or executing the proposed \acrshort{mpt}. $h_{neg}$ denotes that the proposal is expected to be negotiated before the block height $h_{neg}$. Then, the specified executor's \acrshort{tee} $\mathcal{E}^*$ computes $\FuncSty{hash}_{p}$ to be the proposal id $id_p$ and broadcasts a signed $(id_p, p)$ to parties. 

\textit{\textbf{1.2}}: Upon receiving $(id_p, p)$, each $P_i$ interested in the \acrshort{mpt} autonomously responds with a signed acknowledgement $ack_i$ to $E^*$. The $\mathcal{E}^*$ receiving $ack_i$ knows $P_i$'s intent of joining the proposal $id_p$. $\mathcal{E}^*$ keeps collecting $ack_i$ until the acknowledgements match the settlement condition\footnote{Settlement condition of negotiation is flexible, \eg, the number of parties exceeds a specified threshold.} in $\mathcal{P}$. Then, $\mathcal{E}^*$ constructs a settled proposal $p'$ that expands $p$ with the settled parties' addresses $\textit{\textbf{P}}$. Meanwhile, $\mathcal{E}^*$ caches its and parties' coin balances and deducts $q$ collateral from their balance, respectively, ensuring that any involved entity has enough collateral to be punished when it misbehaves. Then, $\mathcal{E}^*$ broadcasts $p'$ to notify the involved parties of the settled proposal.

Otherwise, if $\mathcal{E}^*$ does not collect satisfied acknowledgements, a \textit{challenge-response subprotocol} $Proc_\text{rcha}$ in section~\ref{sec:rcha} will be triggered to identify misbehaviour. We defer the detail in section~\ref{sec:rcha}.

\subsection{Execution phase}
In this phase, $\mathcal{E}^*$ collects plaintext inputs from parties and executes $\mathcal{F}$ to obtain outputs inside \acrshort{tee}. 

\textit{\textbf{2}}: Upon receiving $(id_p, p')$, each party $P_i$ knowing they are involved in the settled proposal $p'$ feeds their inputs (\ie, parameters $x_i$ and old states $s_i$) to $\mathcal{E}^*$.
The $\mathcal{E}^*$ keeps collecting parties' inputs and, especially, reads $\mathcal{F}$-needed old state $\textit{\textbf{s}}$ from $BC$ according to the policy $\mathcal{P}$. If all involved parties' inputs are collected and matched, $\mathcal{E}^*$ executes $\mathcal{F}(\textit{\textbf{s}}, \textit{\textbf{x}})$ to obtain the \acrshort{mpt} outputs, \ie, return values $\textit{\textbf{r}}$ and new states $\textit{\textbf{s}}'$ inside. Then, $\mathcal{E}^*$ goes to the step \textit{\textbf{3.1}}.

Otherwise, if some parties do not submit their inputs as expected, the $Proc_\text{rcha}$ will identify them and punish them. We defer the detail in section~\ref{sec:rcha}.

\subsection{Delivery phase}
This phase adopts an \textit{$\Delta$-fair delivery subprotocol} ($Proc_\text{fdel}$) to reveal the plaintext outputs (\ie, $s'_i, r_i$) to corresponding parties in a $\Delta$-bounded period. The $Proc_\text{fdel}$ proceeds in two steps.

\textit{\textbf{3.1}}
$\mathcal{E}^*$ generates two arrays of symmetric keys $\textit{\textbf{k}}_{s'}, \textit{\textbf{k}}_r$ to computes the commitments of old state and return values $s'_i, r_i$, \ie, $c_{s'_i}, c_{r_i}$, and generates a $proof\gets[H_{\textit{\textbf{c}}_{s}}, H_{\textit{\textbf{c}}_{s'}}]$. The transaction with $proof$ signed by $\mathcal{E}^*$ can prove the \acrshort{mpt}-caused state transition. 
Then, $\mathcal{E}^*$ sends a \textit{Commit} transaction $TX_{cmt}\gets \mathcal{V}.commit(id_p, proof, \textit{\textbf{c}}^*_{s'}, \textit{\textbf{c}}^*_{r}, e_k)$ to commit the outputs on-chain. We note that the published $\textit{\textbf{c}}^*_{s'}, \textit{\textbf{c}}^*_{r}$ do not include the ciphertext of $\textit{\textbf{k}}_s, \textit{\textbf{k}}_r$ so that parties cannot reveal the commitments of $\textit{\textbf{s}}', \textit{\textbf{r}}$. Instead, we require $\mathcal{E}^*$ encrypts the keys with the network key $k_{\mathbfcal{E}}$, where $k_{\mathbfcal{E}}\gets \FuncSty{ECDH}(sk_{\mathbfcal{E}}, pk_{\mathbfcal{E}})$, and attaches the obtained ciphertext $e_k\gets\FuncSty{Enc}_{k_{\mathbfcal{E}}}(\textit{\textbf{k}}_{s'}, \textit{\textbf{k}}_r)$ in $TX_{cmt}$. 
So when $TX_{cmt}$ is confirmed, all $\mathcal{E}\in \mathbfcal{E}$ can read $\textit{\textbf{k}}_{s'}, \textit{\textbf{k}}_r$ on-chain without interacting with each other. Moreover, the $proof$ in $TX_{cmt}$ proves the validity of state transition caused by the \acrshort{mpt} $\mathcal{F}$. $\mathcal{V}$ will validate the $proof$ and lock the on-chain states corresponding to old and new states, which signals the acceptance of the state transition and prevents its corresponding on-chain states from being updated by other concurrent \acrshort{mpt}s before this \acrshort{mpt} completes.

\textit{\textbf{3.2}}: When $TX_{cmt}$ becomes confirmed on-chain, each $E\in \textit{\textbf{E}}$ feeds the $PoP_{cmt}$ (The \acrshort{pop} of the transaction $TX_{cmt}$ which is an enough long and timely block sequence that contains $TX_{cmt}$ to prove $TX_{cmt}$ has been finalized) of $TX_{cmt}$ to its $\mathcal{E}$. Each $\mathcal{E}$ reads key array $\textit{\textbf{k}}_{s'}, \textit{\textbf{k}}_r$ from the $TX_{cmt}$, then sends an transaction $TX_{com}=\mathcal{V}.complete(id_p, [\FuncSty{Enc}_{k_{ie}}(k_{s'_i})], [\FuncSty{Enc}_{k_{ie}}(k_{r_i})])$ to add the ciphertext of $\textit{\textbf{k}}_{s'}, \textit{\textbf{k}}_r$ to $\textit{\textbf{c}}^*_{s'}, \textit{\textbf{c}}^*_r$. The $TX_{com}$ signals the \code{COMPLETED} of this \acrshort{mpt}. 

Here, the delivery fairness is achieved as follows: In \textit{\textbf{3.1}}, each party $P_i$ has received the incomplete output commitments $\textit{\textbf{c}}^*_{s'}, \textit{\textbf{c}}^*_{r}$ but cannot decrypt them without corresponding $k_{s'_i}, k_{r_i}$. In \textit{\textbf{3.2}}, each $\mathcal{E}$ first verifies $PoP_{cmt}$ to ensure that \acrshort{mpt} outputs have been committed on $BC$. Then, each $\mathcal{E}$ can send a $TX_{com}$ to complete the protocol with \code{COMPETED}. Since parties can directly communicate with all executors to obtain $TX_{com}$, they can obtain the $\textit{\textbf{k}}_s, \textit{\textbf{k}}_r$ within the network latency $\Delta$, as long as at least one $E$ honestly respond parties with $TX_{com}$. Otherwise, if $TX_{cmt}$ is rejected by $\mathcal{V}$, any $E$ cannot feed valid $PoP_{cmt}$ to its \acrshort{tee} $\mathcal{E}$. Therefore, no \acrshort{tee} can release $TX_{com}$ to reveal the plaintext outputs or complete the \acrshort{mpt} before $h_{neg}+\tau_{com}$-th block. Therefore, \codename guarantees the $\Delta$-fairness of delivery, where $\Delta$ is the network latency of the blockchain.

\subsection{Challenge-response subprotocol}\label{sec:rcha}

When in any phase one of the honest parties did not receive \acrshort{tee}'s protocol messages as expected, the party can initiate an \textit{challenge-response subprotocol} $Proc_\text{rcha}$. Specifically, it can send a \textit{challengeTEE} transaction $TX_{chaT}$ to challenge the \acrshort{tee} on-chain publicly. The \acrshort{tee} being challenged can only avoid being punished by successfully responding with one of the following transactions:

\begin{itemize}[leftmargin=3mm, parsep=0mm, topsep=1mm, partopsep=0mm]

    \item (i) $TX_{fneg}$: If the $h_{neg}$-th block has not been produced, the \acrshort{tee} $\mathcal{E}^*$ should keep collecting $\textit{\textbf{ack}}$, which are sent by parties from off-chain channels, and $\textit{\textbf{TX}}_{ack}$, which are sent by parties to the blockchain and accepted before the $h_{neg}$-th block. Only if all collected acknowledgement cannot satisfy the settlement condition of \acrshort{mpt} policy $\mathcal{P}$ (If a party $P_i$ send different $ack_i$ by the off-chain channel and the on-chain transaction $TX_{ack_i}$, respectively, the off-chain $ack_i$ will be chosen), $\mathcal{E}^*$ then is allowed to send a $TX_{fneg}$ to fail the proposal on-chain. In all other cases where the $h_{neg}$-th has not been confirmed, or the $\mathcal{E}^*$ has successfully settled the proposal, it's impossible for a \acrshort{tee} to release a $TX_{fneg}$. $TX_{fneg}$ will finish the \acrshort{mpt} as \code{NEGOFAILED}.
    
    \item (ii) $TX_{com}$: If the negotiation phase succeeds and the \acrshort{mpt} completes, a $TX_{com}$ will be sent to the blockchain inherently. $TX_{com}$ will finish the \acrshort{mpt} as \code{COMPLETED}.

    
    \item (iii) $TX_{pnsP}$: If the negotiation phase succeeds, but the $\mathcal{E}^*$ cannot complete the \acrshort{mpt} as expected, both parties and the specified \acrshort{tee}'s executor $E^*$ can be misbehaved entities. Therefore, to avoid being punished in default, $E^*$ should call its $\mathcal{E}^*$ to challenge parties publicly. Specifically, if $\mathcal{E}^*$ does not receive some parties' inputs or match some parties' inputs with their on-chain commitments, $\mathcal{E}^*$ marks these parties as suspicious parties $\textit{\textbf{P}}^*_M$ and returns $\textit{\textbf{P}}^*_M$ to its host $E^*$. The $E^*$ calls $\mathcal{E}^*.challengeParties$ to send a $TX_{chaP}$ to challenge all parties in $\textit{\textbf{P}}^*_M$ on-chain. When $TX_{chaP}$ is confirmed on-chain, honest parties in $\textit{\textbf{P}}^*_M$ are supposed to send a $TX_{resP}$ to publish the ciphertext of their inputs $x_i, s_i$. All published $TX_{resP}$ are required to be confirmed before block height $h_{neg}+\tau_{resP}$. Otherwise, the late $TX_{resP}$ will be regarded as invalid by $\mathcal{E}^*$. Upon the confirmation of the $h_{neg}+\tau_{resP}$-th block, $\mathcal{E}^*$ reads the $PoP_{resP}$ of all $\textit{\textbf{TX}}_{resP}$. If $\mathcal{E}^*$ successfully reads matched inputs of a party $P_i \in \textit{\textbf{P}}^*_M$ from its $TX_{resP_i}$, it removes $P_i$ from $\textit{\textbf{P}}^*_M$. Otherwise, if $PoP_{resP}$ shows that no $TX_{resP_i}$ is published on-chain or the inputs in $TX_{resP_i}$ are still mismatched, $\mathcal{E}^*$ retains $P_i$ in $\textit{\textbf{P}}^*_M$. After that, if $\textit{\textbf{P}}^*_M$ becomes empty, which means all inputs are collected, $\mathcal{E}^*$ goes to the step \textit{\textbf{2}}. Otherwise, if $\textit{\textbf{P}}^*_M$ is not empty, which means the misbehaviour of parties left is confirmed, $\mathcal{E}^*$ marks these parties as $\textit{\textbf{P}}_M$. Then, $\mathcal{E}^*$ sends a $TX_{pnsP}$. $TX_{pnsP}$ calls $punishParties$ to punish the misbehaved parties in finance and signal the \acrshort{mpt} with \code{ABORTED}.
    
\end{itemize}

If the $\mathcal{E}^*$ being challenged by a party either fails (by $TX_{fneg}$), stops (by $TX_{pnsP}$), or completes (by $TX_{com}$) the \acrshort{mpt}, anyone can send a $TX_{pnsT}$ after the $h_{neg}+\tau_{com}$-th block to punish $\mathcal{E}^*$ and signal the \acrshort{mpt} with \code{ABORTED}.  

\section{Implementation} \label{sec:facilities}
\codename is designed to depend on contract-based infrastructure. A service provider of \codename can deploy a contract $\mathcal{V}$ on a legacy $BC$. Then, anyone can interact with the $BC$ and \acrshort{tee}s in $DN$ to transition the states of $BC$ by \acrshort{mpt}s.

\subsection{\codename contract}

We implement the \codename contract in Solidity 0.8.10~\cite{solc}.
As shown in Algorithm~\ref{alg:cloak-service}, $\mathcal{V}$ is constructed by the config of $DN$, \eg, $ad_{\mathbfcal{E}}$, so that parties can authenticate and build secure channels with all $\mathcal{E}\in \mathbfcal{E}$. Moreover, $\mathcal{V}$ provides functions to manage the life cycle of each \acrshort{mpt}. Specifically, a party calls $\mathcal{V}.challengeTEE$ by $TX_{chaT}$ to challenge the specified \acrshort{tee}. and signal the negotiation as \code{NEGOTIATED}. When an \acrshort{mpt} was evaluated, a $\mathcal{E}$ calls $\mathcal{V}.commit$ by $TX_{cmt}$ to validate the state transition and commit the outputs. Finally, a $\mathcal{E}$ calls $\mathcal{V}.complete$ by $TX_{com}$ to release keys' ciphertext and signal the \acrshort{mpt} as \code{COMPLETED}.


\begin{algorithm}[!htbp]
    \footnotesize
    \DontPrintSemicolon
    \LinesNumbered
    \caption{\codename contract $\mathcal{V}$}
    \label{alg:cloak-service}
    \vspace{0.1cm}
    \tcp{This contract is constructed by the network config $ad_{\mathbfcal{E}}$ and a $\acrshort{tee}$ list $\mathbfcal{E}$. $ad_{\mathbfcal{E}}$ is the network account for managing coins deposited by parties. For simplicity, we ignore the register and deposit functions here.
    } 
    \SetKwFunction{FChallengeTEE}{\textit{challengeTEE}}
    \SetKwProg{Fn}{Function}{}{}
    \Fn{\FChallengeTEE{$p$}}{
        \tcp{called by $TX_{chaT}$ from one of parties} 
        $id_p \gets \FuncSty{hash}(p)$ \\
        $\FuncSty{require}(prsls[id_p] = \emptyset)$ \\
        $prsls[id_p].\{q, h_{neg}, \tau_{com}, \mathcal{E}\} \gets p.\{q, h_{neg}\}, \tau_{com}, \mathbfcal{E}[0]$ \\
        
        $prsls[id_p].sta \gets \texttt{PROPOSED}$
    }
    \vspace{0.1cm}
    \SetKwFunction{Facknowledge}{\textit{acknowledge}}
    \SetKwProg{Fn}{Function}{}{}
    \Fn{\Facknowledge{$id_p, \FuncSty{Enc}_{k_{\mathbfcal{E}}}(ack_i)$}}{
        \tcp{called by $TX_{ack}$ from parties} 
        $\FuncSty{require}(BC.\FuncSty{getHeight}() < h_{neg} )$ \\
    }
    \vspace{0.1cm}
    \SetKwFunction{FFailNego}{\textit{failNegotiation}}
    \SetKwProg{Fn}{Function}{}{}
    \Fn{\FFailNego{$id_p$}}{
        \tcp{called by $TX_{fneg}$ from the specified \acrshort{tee}} 
        $\FuncSty{require}(msg.sender = prsls[id_p].\mathcal{E})$ \\
        $prsls[id_p].sta \gets \texttt{NEGOFAILED}$
    }
    \vspace{0.1cm}
    \SetKwFunction{FChallengeParties}{\textit{challengeParties}}
    \SetKwProg{Fn}{Function}{}{}
    \Fn{\FChallengeParties{$id_p, \textit{\textbf{P}}^*_M$}}{
        \tcp{called by $\textit{\textbf{TX}}_{chaP}$ from the specified \acrshort{tee}} 
    }
    \vspace{0.1cm}
    \SetKwFunction{FResponse}{\textit{partyResponse}}
    \SetKwProg{Fn}{Function}{}{}
    \Fn{\FResponse{$id_p, \FuncSty{Enc}_{k_{\mathbfcal{E}}}(in)$}}{
        \tcp{called by $\textit{\textbf{TX}}_{resP}$ from parties} 
        $\FuncSty{require}(BC.\FuncSty{getHeight}() < h_{neg} + \tau_{resP} )$ \\
    }
    \vspace{0.1cm}
    \SetKwFunction{FPunish}{\textit{punishParties}}
    \SetKwProg{Fn}{Function}{}{}
    \Fn{\FPunish{$id_p, \textit{\textbf{P}}_M$}}{
        \tcp{called by $TX_{pnsP}$ from the specified \acrshort{tee}}
        $\FuncSty{require}(msg.sender = prsls[id_p].\mathcal{E})$ \\
        \tcp{update coins for punishment}
        \textbf{for} $P_i \in \textit{\textbf{P}}_M$ \textbf{do} \\
            \qquad$coins[P_i] \gets coins[P_i] -q$ \\
        $prsls[id_p].sta \gets \texttt{ABORTED}$
    }
    \vspace{0.1cm}
    \SetKwFunction{FCommit}{\textit{commit}}
    \SetKwProg{Fn}{Function}{}{}
    \Fn{\FCommit{$id_p, proof, \textit{\textbf{c}}^*_{s'}, \textit{\textbf{c}}^*_{r}, e_k$}}{
        \tcp{called by $TX_{cmt}$ from the specified \acrshort{tee}} 
        $\FuncSty{require}(msg.sender = prsls[id_p].\mathcal{E})$ \\
        $\FuncSty{require}(\FuncSty{verify}(proof,~H_{\textit{\textbf{c}}_s}))$ \tcp{match old states}
    }
    \vspace{0.1cm}
    \SetKwFunction{FComplete}{\textit{complete}}
    \SetKwProg{Fn}{Function}{}{}
    \Fn{\FComplete{$id_p, [\FuncSty{Enc}_{k_{ie}}(k_{s'_i})|_{1..n}],~[\FuncSty{Enc}_{k_{ie}}(k_{r_i})|_{1..n}]$}}{
        \tcp{called by $TX_{com}$ from \textbf{any registered \acrshort{tee}}} 
        $\FuncSty{require}(msg.sender \in \mathbfcal{E})$ \\
        $H_{\textit{\textbf{c}}_s} \gets proof.H_{\textit{\textbf{c}}_{s'}}$ \tcp{set new states}
        $prsls[id_p].sta \gets \texttt{COMPLETED}$ \\
    }
    \vspace{0.1cm}
    \SetKwFunction{FSettle}{\textit{punishTEE}}
    \SetKwProg{Fn}{Function}{}{}
    \Fn{\FSettle{$id_p$}}{
        \tcp{called by $TX_{pnsT}$ from anyone} 
        $\FuncSty{require}(prsls[id_p] \neq \emptyset~\textbf{and}~BC.\FuncSty{getHeight}()>h_{neg} + \tau_{com})$ \\
        $\FuncSty{require}(prsls[id_p].sta \notin \{\texttt{NEGOFAILED}, \texttt{ABORTED}, \texttt{COMPLETED}\})$ \\
        $coins[prsls[id_p].\mathcal{E}] \gets coins[prsls[id_p].\mathcal{E}] - q$ \\
        $prsls[id_p].sta \gets \texttt{ABORTED}$
    }
    \vspace{0.1cm}
\end{algorithm}

    
    


\subsection{\codename network}
To construct the $DN$, we instantiate each \acrshort{tee} $\mathcal{E}$ (Algorithm~\ref{alg:cloak-enclave}) based on SGX~\cite{costan2016intel}. 
Anyone with a \acrshort{tee} device can instantiate a $\mathcal{E}$ (Algorithm~\ref{alg:cloak-enclave}) to become a executor $E$. The first $\mathcal{E}$ generates the network account $(sk_{\mathbfcal{E}}, pk_{\mathbfcal{E}}, ad_{\mathbfcal{E}})$ to initialize a network $DN$. Then, other $\mathcal{E}$ must be attested by one of $\mathcal{E}$ in the $DN$ to join the $DN$ and obtain the network key and account. 

\begin{algorithm}[!h]
    \footnotesize
    \DontPrintSemicolon
    \LinesNumbered
    \caption{\codename enclave program ($\mathcal{E}$)}
    \label{alg:cloak-enclave}
    \vspace{0.1cm}
    \tcp{For simplicity, we assume each $\mathcal{E}$ has obtained the network config and cached the balances of parties' coins by synchronization. The config includes a secure parameter $\kappa$, a checkpoint $b_{cp}$ of $BC$, and the network account $(sk_{\mathbfcal{E}}, pk_{\mathbfcal{E}}, ad_{\mathbfcal{E}})$.
    }
    \SetKwFunction{FGenIDp}{\textit{generateIDp}}
    \SetKwProg{Fn}{Procedure}{}{}
    \Fn{\FGenIDp{$p$}}{
        \tcp{check this is the specified \acrshort{tee}}
        \textbf{if} $ self\neq BC.\mathbfcal{E}[0] $ \textbf{then} abort \\
        $id_p \gets \FuncSty{hash}(p)$ \\
        \Return $( id_p, p )$
    }
    \vspace{0.1cm}
    \SetKwFunction{Fnegotiate}{\textit{negotiate}}
    \SetKwProg{Fn}{Procedure}{}{}
    \Fn{\Fnegotiate{$id_p, \textit{\textbf{ack}}$}}{
        \textbf{if} $ status = \texttt{NEGOTIATED} $ \textbf{then} \Return $(id_p, p')$ \\
        \textbf{if} $ status \neq \emptyset $ \textbf{or} $\FuncSty{conform}(\textit{\textbf{ack}}, \mathcal{P})\neq1$ \\
          \quad\textbf{or} $ cacheCoins[self] - q < 0 $\\ 
          \quad\textbf{or} $\exists P_i \in \textit{\textbf{P}},  cacheCoins[P_i]- q <0$  \textbf{then} abort \\
        $p', status \gets (p.\{H_{\mathcal{F}},H_{\mathcal{P}},q, h_{neg}\}), ~\texttt{NEGOTIATED} $ \\
        \Return $(id_p, p')$
    }
    \vspace{0.1cm}
    \SetKwFunction{FfailNego}{\textit{failNegotiation}}
    \SetKwProg{Fn}{Procedure}{}{}
    \Fn{\FfailNego{$id_p, TX_{chaT}, PoP_{chaT}$}}{
        \textbf{if} $ status \neq \emptyset $ \textbf{or} $\FuncSty{veriPoP}(b_{cp}, PoP_{chaT}, TX_{chaT})\neq 1$ \textbf{then} abort \\
        \textbf{if} $PoP_{chaT}.\FuncSty{getComfHeight}() > p.h_{neg}$ \textbf{then} \\
            \qquad $\textit{\textbf{TX}}_{ack} \gets$ all $PoP_{chaT}.\textit{\textbf{TX}}_{ack_i}$ before $p.h_{neg}$ \\
            \qquad\textit{\textbf{ack}} $\gets \textit{\textbf{ack}} \cup \textit{\textbf{TX}}_{ack}.\textit{\textbf{ack}}$ \\
        \textbf{if} $\FuncSty{conform}(\textit{\textbf{ack}}, \mathcal{P})=1$ \textbf{then} abort\\
        \Return $\textit{\textbf{TX}}_{fneg}(id_p)$  \\
    }
    \vspace{0.1cm}
    \SetKwFunction{FEexecute}{\textit{execute}}
    \SetKwProg{Fn}{Procedure}{}{}
    \Fn{\FEexecute{$id_p, \textit{\textbf{in}}, PoP_{s}$}}{
        \textbf{if} $status \neq \texttt{NEGOTIATED}$ \textbf{then} abort \\
        $\textit{\textbf{P}}^*_M \gets \textit{\textbf{P}}$ \\
        \textbf{for} $x_i, k_{x_i}$~\textbf{in}~$\textit{\textbf{in}}.\{\textit{\textbf{x}},~\textit{\textbf{k}}_x\}$ \\
                \qquad$\textit{\textbf{P}}^*_M \gets \textit{\textbf{P}}^*_M \backslash \{P_i\}$ \\
        \textbf{if} $|\textit{\textbf{P}}^*_M|>0$ \textbf{then} \Return $(id_p, \textit{\textbf{P}}^*_M)$  \\
        \tcp{evaluates $\mathcal{F}(x)$ on states $s$} 
        $\textit{\textbf{s}}', \textit{\textbf{r}} \gets \mathcal{F}(PoP_{s}.\textit{\textbf{s}}, \textit{\textbf{x}})$ \\
        $b_{cp}\gets PoP_{s}.\FuncSty{getLastComfBlock}()$ \\
        $status \gets \texttt{EXECUTED}$ \\
    }
    \vspace{0.1cm}
    \SetKwFunction{Fcommit}{\textit{commit}}
    \SetKwProg{Fn}{Procedure}{}{}
    \Fn{\Fcommit{$id_p$}}{
        \textbf{if} $status \neq \texttt{EXECUTED}$ \textbf{then} abort \\
        $\textit{\textbf{k}}_{s'}, \textit{\textbf{k}}_r \gets Gen(1^\kappa)$ \\
        $c_{s'_i} \gets [\FuncSty{Enc}_{k_{s'_i}}(s'_i),\FuncSty{Enc}_{k_{ie}}(k_{s'_i}),P_i]$ \\
        $proof \gets [PoP_{s}.H_{\textit{\textbf{c}}_s}, H_{\textit{\textbf{c}}_{s'}}]$ \\
        $c^*_{s'_i},c^*_{r_i} \gets [\FuncSty{Enc}_{k_{s'_i}}(s'_i),0,P_i],~ [\FuncSty{Enc}_{k_{r_i}}(r_i),0,P_i]$ \\
        \Return $TX_{cmt}(id_p, proof, \textit{\textbf{c}}^*_{s'}, \textit{\textbf{c}}^*_r, e_k)$
    }
    \vspace{0.1cm}
    \SetKwFunction{FEchallenge}{\textit{challengeParties}}
    \SetKwProg{Fn}{Procedure}{}{}
    \Fn{\FEchallenge{$\textit{\textbf{P}}^*_M$}}{
        \textbf{if} $status \neq \texttt{NEGOTIATED}$ \textbf{then} abort \\
        \textbf{if} $|\textit{\textbf{P}}^*_M|>0$ \textbf{then} \\
            \qquad\Return $\textit{\textbf{TX}}_{chaP}(id_p, \textit{\textbf{P}}^*_M)$  \\
    }
    \vspace{0.1cm}
    \SetKwFunction{FEpunish}{\textit{punishParties}}
    \SetKwProg{Fn}{Procedure}{}{}
    \Fn{\FEpunish{$\textit{\textbf{TX}}_{chaP}, \textit{\textbf{TX}}_{resP}, PoP_{resP}$}}{
        \textbf{if} $status \neq \texttt{NEGOTIATED}$ \textbf{or} $\FuncSty{veriPoP}(b_{cp}, \textit{\textbf{TX}}_{chaP}, PoP_{resP})\neq 1$ \textbf{then} abort \\
        $\textit{\textbf{P}}_M \gets \textit{\textbf{P}}^*_M$ \\
        \textbf{for} $P_i \in \textit{\textbf{P}}^*_M$ \textbf{do} \\
            \qquad\textbf{if} $x_i, k_{x_i} \gets TX_{resP_i}.\{x_i, k_{x_i}\}$ \textbf{then} \\
                \qquad\qquad$\textit{\textbf{P}}_M \gets \textit{\textbf{P}}_M \backslash \{P_i\}$ \\
        \textbf{if} $|\textit{\textbf{P}}_M|>0$ \textbf{then}  \\
            \qquad\Return $TX_{pnsP}(id_p, \textit{\textbf{P}}_M)$ \\
    }
    \vspace{0.1cm}
    \SetKwFunction{FDcomplete}{\textit{complete}}
    \SetKwProg{Fn}{Procedure}{}{}
    \Fn{\FDcomplete{$TX_{cmt}, PoP_{cmt}$}}{
        \textbf{if} $ status\neq \texttt{NEGOTIATED}$ \textbf{or} $\FuncSty{veriPoP}(b_{cp}, TX_{cmt}, PoP_{cmt})\neq 1$ \textbf{then} abort \\
        $status \gets \texttt{COMPLETED} $ \\
        \Return $TX_{com}(id_p, [\FuncSty{Enc}_{k_{ie}}(k_{s'_i})|_{i\in [n]}],~[\FuncSty{Enc}_{k_{ie}}(k_{r_i})|_{i\in [n]}])$
    }
    \vspace{0.1cm}
\end{algorithm}

To evaluate \acrshort{mpt}, we express $\mathcal{F}$ in Solidity 0.8.10~\cite{solc} and port EVM~\cite{evm} into SGX. $\mathcal{P}$ is expressed in JSON. $\mathcal{P}$ is introduced to specify the parameters, states to read and write, and return values of $\mathcal{F}$, which is for \acrshort{tee} to know the I/O of the \acrshort{mpt}.
The hash of both $\mathcal{F}$ and $\mathcal{P}$ are registered and updated on $BC$ , while their codes are provided by the \acrshort{mpt}' developers/initiators and cached by $\mathbfcal{E}$. Admittedly, $\mathcal{P}$ is now pre-specified thus restricting that the I/O of $\mathcal{F}$ should be statically identified. However, this problem could solved by hooking EVM's \code{sstore} and \code{sload} instructions~\cite{secondstate'20}, and we leave it for future work. 

\section{Security Analysis} \label{sec:security}

\subsection{Assumption reliability}
Our assumption that \acrshort{tee}’s confidentiality and attestable integrity hold is still practical now. While attacks against SGX, \eg, memory-corruption attacks and side-channel attacks, keep coming out, the community has developed efficient software-based~\cite{gruss2017strong, shih2017t, Lang2022MoLEMO} and hardware-based countermeasures~\cite{noorman2013sancus, costan2016sanctum}. So far, most of existing attacks against SGX are either function-limited~\cite{Biondo2018codereuse, Brasser2017cacheattack}, solved, or patched~\cite{Bulck2018foreshadow, l1tf}. For some very recent and considerable attacks like xAPIC and MMIO, they are also mitigated in Dec. 22 and will be solved in Jan. 23~\cite{sgxfail}.

\subsection{Protocol security}
Informally, we claim that the following theorem holds. We leave the formal security property definition and corresponding game theory-based proof in Appendix~\ref{sec:security-proof}. Limited by space, here we will briefly outline the idea of how we prove \textit{financial fairness}, and \textit{delivery fairness}.


\begin{theorem-box}
    \begin{informal-theorem}[Informal statement] \textit{The protocol $\pi_{\codename}$ satisfies \textbf{correctness}, \textbf{confidentiality}, \textbf{public verifiability}, \textbf{data availability}, \textbf{financial fairness}, \textbf{delivery fairness}, and \textbf{delivery atomicity}}
    \end{informal-theorem}
\end{theorem-box}

To prove \codename holds \textit{financial fairness}, we prove that there are only three possible statuses of an \acrshort{mpt}, \ie, $\emptyset$ (negotiation not started or gets failed), $\code{ABORTED}$ (negotiation succeeded, but the \acrshort{mpt} did not complete as expected) and $\code{COMPLETED}$ (the \acrshort{mpt} complete as expected). Then, we exhaustively prove that parties' balance will stay fair in any of the three statuses: i) if the status of an \acrshort{mpt} stays at $\emptyset$, all entities' balances would have no change; ii) if an \acrshort{mpt}'s status is $\code{ABORTED}$, then either some parties misbehaved and were punished, or the specified \acrshort{tee} executor misbehaved and were punished; iii) if an \acrshort{mpt}'s status becomes $\code{COMPLETED}$, the \acrshort{mpt} succeeds, and all entities' balances would have no change.

To prove the \textit{delivery fairness} being held, we utilize the ideal availability of blockchain and the assumption that all-but-one \acrshort{tee} executors are Byzantine. Specifically, to release outputs, the $TX_{cmt}$, which contains data ciphertext and the ciphertext of their corresponding keys, must have been published on the blockchain. Therefore, if each party communicate with all \acrshort{tee} executors directly and at least one \acrshort{tee} node is honest, all parties can obtain their corresponding outputs in the $\Delta$-bounded period. The $\Delta$ equals to the message delivery upper bound of the (semi-)synchronous network among parties and \acrshort{tee} nodes.

\section{Evaluation} \label{sec:evaluation}
\myparagraph{Methodology and setup}
To evaluate the effectiveness of \codename, we propose 3 research questions.
\begin{itemize}[leftmargin=3mm, parsep=0mm, topsep=1mm, partopsep=0mm]
    \item \textbf{Q1}: Can \codename capably serve real-world \acrshort{mpt}s?
    \item \textbf{Q2}: What is the cost of enabling \acrshort{mpt}s on a blockchain?
    \item \textbf{Q3}: What is the cost of evaluating \acrshort{mpt}s using \codename?
\end{itemize}

The experiment is based on a server with Ubuntu 18.04, 32G memory, and 2.2GHz Intel(R) Xeon(R) Silver 4114 CPU. The memory used by \acrshort{tee} is set up to 200M.

\myparagraph{Answering Q1}
We evaluate \codename on 5 contracts which involve 10 \acrshort{mpt}s in different scenarios. All them are in Solidity and the number of parties they involved varies from 2-11.

\textit{SupplyChain} is a contract allowing suppliers to negotiate and privacy-preservedly bids off-chain, and commit the evaluation with their new balances on-chain. It has 39 LOC and contains one \acrshort{mpt}.

\textit{Scores} is a contract allowing students to join and get mean scores off-chain and commit
the evaluation on-chain. It has 95 LOC and contains one \acrshort{mpt}.

\textit{ERC20Token} is a contract allowing accounts to pair and transfer without revealing balances
off-chain, and commit the evaluation with new balances on-chain. It has 55 LOC and contains three \acrshort{mpt}s.

\textit{YunDou} is a fine-tuned ERC20 token contract with co-managed accounts where account managers self-selectly vote to transfer tokens without revealing the votes. It has 105 LOC and contains three \acrshort{mpt}s.

\textit{Oracle} is a Oracle contract that allows parties to negotiate to join then jointly and verifiably generate random numbers. It has 60 LOC and contains three \acrshort{mpt}s.

\myparagraph{Answering Q2}
Table~\ref{tab:gas-cost} shows the gas cost of all methods of $\mathcal{V}$ in different phases. To answer Q2, here we focus on the initialization and global setup phase.

\begin{table}[!htbp]
  \centering
  \caption{\textbf{On-chain cost of challenge-response submission phase.} \small{For each \acrshort{mpt}, we assume all partied involved are challenged}}
  \vspace{-0.2cm}
  \label{tab:gas-cost}
  \setlength{\tabcolsep}{3mm}{
      \footnotesize
      \begin{tabular}{lcc}
        \toprule
        \textbf{Phase} & \textbf{TX}  & \textbf{Gas cost} \\
        \midrule
        \multirow{2}{*}{Global setup} & \textit{register} ($TX_{reg_i}$) & 127068 \\
        & \textit{deposit} ($TX_{dep_i}$) & 42325 \\
        \cmidrule{2-3} \multirow{2}{*}{\acrshort{mpt}}
        & \textit{commit} ($TX_{cmt}$) & 104568 \\ 
        & \textit{complete} ($TX_{com}$) & 110570 \\ 
        \cmidrule{2-3} \multirow{7}{*}{$Proc_\text{rcha}$}
        & \textit{challengeTEE} ($TX_{chaT}$) & 131762 \\ 
        & \textit{acknowledge} ($TX_{ack_i}$) & 26999 \\ 
        & \textit{failNegotiation} ($TX_{fneg}$) & 30563 \\ 
        & \textit{challengeParties} ($TX_{chaP}$) & 33786 \\ 
        & \textit{partyResponse} ($TX_{resP_i}$) & 34313 \\ 
        & \textit{punishParties} ($TX_{pnsP}$) & 45518 \\
        & \textit{punishTEE} ($TX_{pnsT}$) & 53254 \\
        \cmidrule{1-3} 
        & \textit{DeFi: ERC20: Transfer} & 65000 \\
        & \textit{DeFi: Uniswap V3: Swap} & 184523 \\
        & \textit{DeFi: Balancer: Swap} & 196625 \\
        & \textit{NFT: OpenSea: Sale} & 71645 \\ 
        & \textit{NFT: LooksRare: Sale} & 326897 \\
        \bottomrule
      \end{tabular}
  }
\end{table}

\textit{Gas cost of initialization}. 
It costs 4.9M gas to deploy $\mathcal{V}$ to enable \codename on a blockchain. This cost is only once paid by \codename service provider, thereby is irrelevant.

\textit{Gas cost of global setup}. 
A party pays 12.7k to \textit{register} its public key and 4.2k gas to \textit{deposit} coins. This setup happens once for each party, thus being acceptable. 

\myparagraph{Answering Q3} We analyze the gas and off-chain cost for evaluating each \acrshort{mpt}, respectively. Especially, we compare the gas cost of \codename with the most related \acrshort{mpt}-oriented work, Fastkitten~\cite{FastKitten'19} and Cloak~\cite{ren2022Cloak}.

\begin{figure}[!htbp]
    \vspace{-0.3cm} 
    \setlength{\belowcaptionskip}{-1cm} 
    \centering
    \includegraphics[width=8cm]{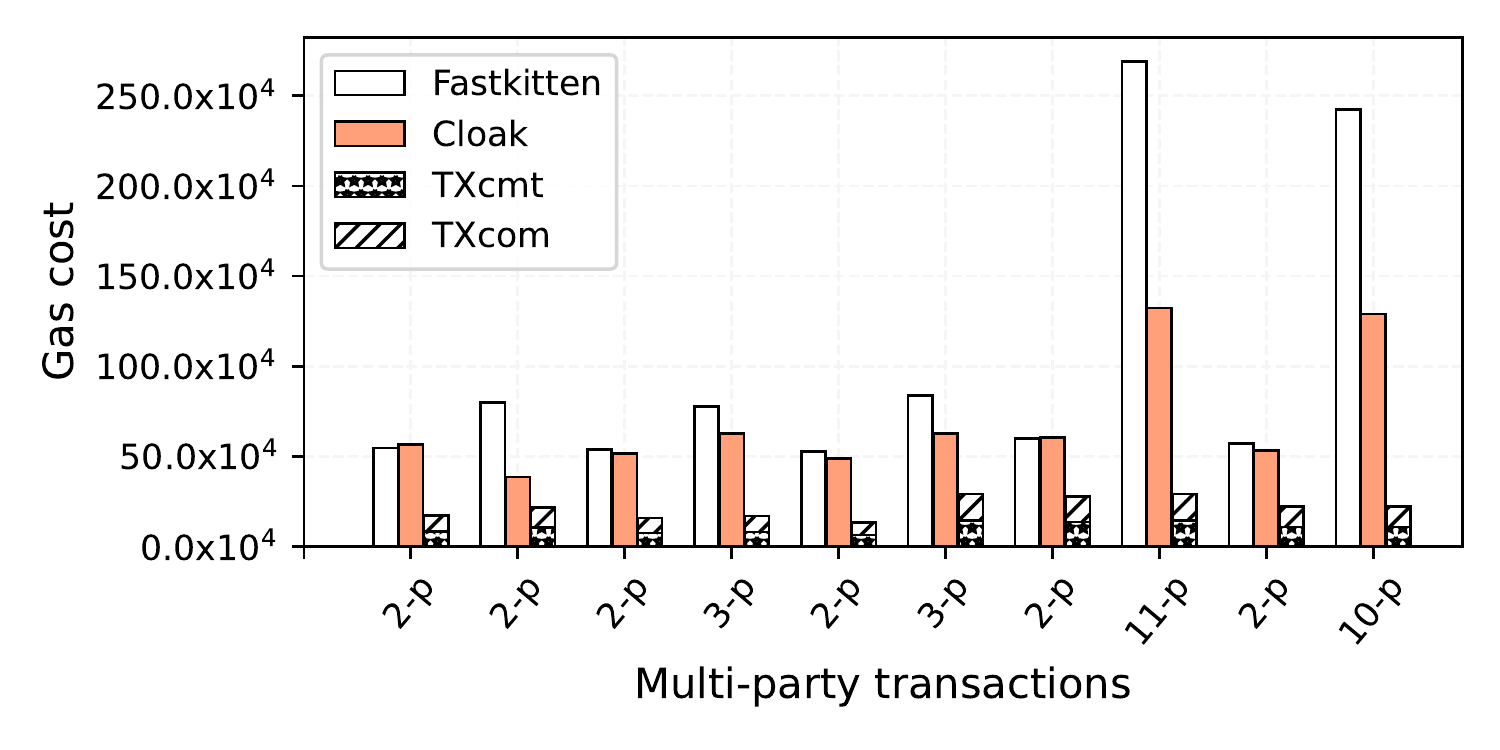}
    \caption{\textbf{The gas cost of \codename}. \small{``Fastkitten" refers to the gas cost sum of $n+1$ transactions for each \acrshort{mpt}. Here we adapt the protocol of Faskkitten to Ethereum. ``Cloak" refers to the gas cost sum of its $2$ transactions for each \acrshort{mpt}. ``TXcmt" and ``TXcom" refers to gas cost of $TX_{cmt}, TX_{com}$} in $\pi_{\codename}$, respectively.
    }
    \label{fig:gas-cost}
\end{figure}

\textit{On-chain cost of \acrshort{mpt}s}.
Figure~\ref{fig:gas-cost} shows the gas cost of each \acrshort{mpt}. Overall, \codename reduces gas by 72.5\% against Fastkitten. Specifically, for six 2-party \acrshort{mpt}s, \codename costs 0.27-0.46X gas. For two 3-party and two 10/11-party \acrshort{mpt}s, the gas significantly reduces to 0.22-0.25X and 0.09-0.11X, respectively. For Cloak, the cost of \codename decreases by 65.6\% in average. Specifically, \codename costs 0.27-0.56X gas against Cloak in 2/3-party \acrshort{mpt}s, while just 0.17-0.22X gas in 10/11-party \acrshort{mpt}s. Therefore, \codename enables a more secure \acrshort{mpt}s with lower on-chain cost. The on-chain cost not only surpasses Cloak, but is comparable to typical single-party transactions, \eg, NFT sale and ERC20 swap, on Ethereum. Moreover, as the number of parties growing, the cost superiority of \codename improves.

\textit{Off-chain cost of \acrshort{mpt}s}.
All 10 \acrshort{mpt}s complete in constant 2 transactions. Specifically, the negotiation, execution, and delivery phases cost 0.21-0.58s, 0.39-1.15s, and 0.30-0.77s, respectively, which can be ignored. 



\section{Optimization and fine-tuning} \label{sec:discusion}

\subsection{Improve the scalability of \codename}

\subsubsection{Reduce gas cost in optimistic cases}\label{optimistic-cost-optimization} 
Recall that serving an \acrshort{mpt} in optimistic scenarios only involves 2 transactions, $TX_{cmt}$ and $TX_{com}$. Therefore, to serve a $n$-party \acrshort{mpt} without adversary, \codename needs to send $O(1)$ transactions. We note that we can adopt the following measures to furthermore reduce the optimistic cost of \codename.

\myparagraph{batch processing}
According to the height of the blockchain, we can split the execution of \acrshort{mpt} to different slots. In each slot, \codename handles $\lambda$ \acrshort{mpt}s ($\lambda\geq1$) and sends only two transactions, \ie $TX_{cmt}, TX_{com}$, to finish all \acrshort{mpt}s in the slot in a batch. This way, it can reduce the complexity to $O(1/\lambda)$ without sacrificing the security or changing the adversary model. 

\myparagraph{making trade-off}
We note that by intentionally sacrificing some of our system goals, \codename can furthermore reduce its on-chain cost. First, we can drop \textit{data availability} to delete the last transaction $TX_{com}$. Specifically, in the delivery phase, \acrshort{tee}s will first send $TX_{cmt}$ to commit outputs on-chain. If the $proof$ in $TX_{cmt}$ passes, $\mathcal{V}$ will accept the state transition immediately. Then, upon $TX_{cmt}$ being accepted and confirmed, \acrshort{tee}s will release the keys of the output ciphertext in $TX_{cmt}$ to parties by off-chain channels, rather than sending a $TX_{com}$. Consequently, the required transactions of \codename reduce to only 1, \ie, $TX_{cmt}$. However, in this variant, parties need to keep all received keys to access their plaintext states. Second, we can furthermore drop \textit{delivery atomicity} and \textit{delivery fairness} to delete $TX_{cmt}$, meaning that no transactions are required in the optimistic case. Specifically, \acrshort{mpt} involves reading on-chain inputs. If we delete $TX_{cmt}$, when the specified \acrshort{tee} obtains outputs, the blockchain has no change to ensure that old states that \acrshort{mpt} read have not been mutated. This way, the \acrshort{mpt} outputs that \acrshort{tee} regard as valid cannot be accepted by the blockchain, breaking the atomicity. Moreover, as we cannot utilize the $TX_{cmt}$ to ensure that output ciphertext can be ideally delivered to all \acrshort{tee}s, \textit{delivery fairness} is broken.

\subsubsection{Reduce gas cost in pessimistic cases}\label{pessimistic-cost-optimization}
In the pessimistic scenarios, the \textit{challenge-response protocol} ($Proc_\text{rcha}$) will be triggered. In the protocol, each party being challenged on-chain has to respond with their acknowledgements or inputs independently. We can introduce an off-chain third-party service to collect parties' responses and publish an aggregated $TX_{resP}$ to the blockchain. In this, way, even though a $Proc_\text{rcha}$ is being triggered, the on-chain transaction complexity is still $O(1)$. And combining with the batch processing technique of \acrshort{mpt}, the complexity of $Proc_\text{rcha}$ can furthermore reduce to $O(1/m)$, where $m$ is the number of \acrshort{mpt}s in a batch.

\subsubsection{Reduce storage cost} To minimize the trust of off-chain \acrshort{tee} network, \codename stores parties' privacy-preserved data on blockchain and ensure the plaintext of the stored data are still accessible to parties even without \codename. This sounds indicating a heavy storage cost. However, as we demonstrated in Section~\ref{sec:evaluation}, the storage cost is acceptable. Actually, storing off-chain states on-chain as \code{calldata} has been well-adopted in Ethereum Rollup projects~\cite{oprollup, zkrollup}. Moreover, reducing the storage cost is also a main issue of Ethereum 2.0. Specifically, Ethereum propose to reduce the gas cost of \code{calldata} from 16 to 3, which means a 81\% decrease~\cite{eip-4488}. Furthermore, Ethereum 2.0 will introduce \code{blob}~\cite{eip-4844}, a new storage mechanism which allows different Ethereum Layer-2 projects to cheaply store all their transactions and states on the Beacon chain. Therefore, the design of \codename strongly match the need and tendency of Ethereum.

\subsection{Improve the availability of \codename}
An industry \textit{tee} service usually has a robust error-handling mechanism and is DDoS-resistant. Therefore, we practically assume that the service provided by the specified honest \acrshort{tee} executor is highly available. However, it does mean we cannot further improve the availability of \codename. For example, \codename can adopt a similar availability enhancement mechanism as in POSE~\cite{FrassettoNDSS22POSE}. Specifically, every time the specified \acrshort{tee} executor changes its local state, it should synchronize the state updates to all other registered \acrshort{tee}s and collect their signatures in off-chain channels to carry on the next state transfer. If the specified \acrshort{tee} is not available off-chain, parties can publicly change it on-chain. If the unavailability of the specified \acrshort{tee} is because that other \acrshort{tee} executors do not respond with signatures as expected, the specified \acrshort{tee} can publicly challenge other unavailable \acrshort{tee}s on the blockchain. Finally, if the on-chain challenge-response mechanism finally punishes the specified \acrshort{tee}, it will be kicked out, and the next \acrshort{tee} in the registered list will be specified to serve \acrshort{mpt}s. As a result, in an optimistic scenario, \ie, all other \acrshort{tee}s honestly respond with their signatures, \codename will not lose its off-chain states if at least one \acrshort{tee} is available. In a word, we stress that improving the availability of \acrshort{tee} network is an orthogonal field with \codename, and \codename can combine with the related work~\cite{FrassettoNDSS22POSE} to further improve its availability.

\section{Conclusion} \label{sec:conclusion}
In this paper, we develop a novel framework, \codename, which can support \acrshort{mpt}-enabled off-chain contract execution on legacy blockchains by using a \acrshort{tee} network.
\codename features maximising the security of \acrshort{mpt} and minimising the gas cost and the network's trust. 
Comparing with the SOTA, Cloak~\cite{ren2022Cloak}, \codename not only realizes all security properties the SOTA claimed but also additionally achieves data availability, delivery fairness, and delivery atomicity. To our knowledge, \codename achieves the most general and secure \acrshort{mpt}. 
Meanwhile, it assumes at least one party and executor are honest, which is also one of the weakest assumptions compared to related work. 
Moreover, according to our evaluation, \codename reduces the gas cost of the SOTA by 65.6\%, and the superiority of \codename increases as the number of parties grows.




\bibliographystyle{IEEEtran}
\bibliography{bibliography}

\begin{thebibliography}{10}
\providecommand{\url}[1]{#1}
\csname url@samestyle\endcsname
\providecommand{\newblock}{\relax}
\providecommand{\bibinfo}[2]{#2}
\providecommand{\BIBentrySTDinterwordspacing}{\spaceskip=0pt\relax}
\providecommand{\BIBentryALTinterwordstretchfactor}{4}
\providecommand{\BIBentryALTinterwordspacing}{\spaceskip=\fontdimen2\font plus
\BIBentryALTinterwordstretchfactor\fontdimen3\font minus
  \fontdimen4\font\relax}
\providecommand{\BIBforeignlanguage}[2]{{%
\expandafter\ifx\csname l@#1\endcsname\relax
\typeout{** WARNING: IEEEtran.bst: No hyphenation pattern has been}%
\typeout{** loaded for the language `#1'. Using the pattern for}%
\typeout{** the default language instead.}%
\else
\language=\csname l@#1\endcsname
\fi
#2}}
\providecommand{\BIBdecl}{\relax}
\BIBdecl

\bibitem{nakamoto2008bitcoin}
S.~Nakamoto, ``Bitcoin: A peer-to-peer electronic cash system,''
  \emph{Decentralized Business Review}, p. 21260, 2008.

\bibitem{wood2014ethereum}
G.~Wood \emph{et~al.}, ``Ethereum: A secure decentralised generalised
  transaction ledger,'' \emph{Ethereum project yellow paper}, 2014.

\bibitem{Hawk:SP2016}
A.~Kosba, A.~Miller, E.~Shi, Z.~Wen, and C.~Papamanthou, ``{Hawk: The
  Blockchain Model of Cryptography and Privacy-Preserving Smart Contracts},''
  \emph{2016 IEEE Symposium on Security and Privacy (SP)}, pp. 839--858, 2016.

\bibitem{Sinha2020LucidiTEEAT}
R.~Sinha, ``Luciditee: A tee-blockchain system for policy-compliant multiparty
  computation with fairness,'' 2020.

\bibitem{GovinICBC22:Rialto}
K.~Govindarajan, D.~Vinayagamurthy, P.~Jayachandran, and C.~Rebeiro,
  ``Privacy-preserving decentralized exchange marketplaces,'' in \emph{2022
  IEEE International Conference on Blockchain and Cryptocurrency (ICBC)}, 2022,
  pp. 1--9.

\bibitem{MassacciSP18:FuturesMEX}
F.~Massacci, C.~N. Ngo, J.~Nie, D.~Venturi, and J.~Williams, ``Futuresmex:
  Secure, distributed futures market exchange,'' in \emph{2018 IEEE Symposium
  on Security and Privacy (SP)}, 2018, pp. 335--353.

\bibitem{ren2021CloakDemo}
Q.~Ren, H.~Liu, Y.~Li, and H.~Lei, ``Demo: Cloak: A framework for development
  of confidential blockchain smart contracts,'' in \emph{2021 IEEE 41st
  International Conference on Distributed Computing Systems (ICDCS)}, 2021, pp.
  1102--1105.

\bibitem{ren2022Cloak}
Q.~Ren, Y.~Wu, H.~Liu, Y.~Li, A.~Victor, H.~Lei, L.~Wang, and B.~Chen, ``Cloak:
  Transitioning states on legacy blockchains using secure and publicly
  verifiable off-chain multi-party computation,'' in \emph{Proceedings of the
  38th Annual Computer Security Applications Conference}, 2022, pp. 117--131.

\bibitem{CarstenSCN'14PAMPC}
C.~Baum, I.~Damg{\aa}rd, and C.~Orlandi, ``Publicly auditable secure
  multi-party computation,'' in \emph{Security and Cryptography for Networks},
  M.~Abdalla and R.~De~Prisco, Eds.\hskip 1em plus 0.5em minus 0.4em\relax
  Cham: Springer International Publishing, 2014, pp. 175--196.

\bibitem{DanIACR19ZKP}
\BIBentryALTinterwordspacing
D.~Boneh, E.~Boyle, H.~Corrigan-Gibbs, N.~Gilboa, and Y.~Ishai,
  ``Zero-knowledge proofs on secret-shared data via fully linear pcps,''
  Cryptology ePrint Archive, Paper 2019/188, 2019,
  \url{https://eprint.iacr.org/2019/188}. [Online]. Available:
  \url{https://eprint.iacr.org/2019/188}
\BIBentrySTDinterwordspacing

\bibitem{cui2021mpc}
H.~Cui, K.~Zhang, Y.~Chen, Z.~Liu, and Y.~Yu, ``Mpc-in-multi-heads: A
  multi-prover zero-knowledge proof system,'' in \emph{European Symposium on
  Research in Computer Security}.\hskip 1em plus 0.5em minus 0.4em\relax
  Springer, 2021, pp. 332--351.

\bibitem{SamuelSP2022ZeeStar}
S.~Steffen, B.~Bichsel, R.~Baumgartner, and M.~Vechev, ``Zeestar: Private smart
  contracts by homomorphic encryption and zero-knowledge proofs,'' in
  \emph{2022 IEEE Symposium on Security and Privacy (SP)}, 2022, pp. 179--197.

\bibitem{FastKitten'19}
\BIBentryALTinterwordspacing
P.~Das, L.~Eckey, T.~Frassetto, D.~Gens, K.~Host{\'a}kov{\'a}, P.~Jauernig,
  S.~Faust, and A.-R. Sadeghi, ``Fastkitten: Practical smart contracts on
  bitcoin,'' in \emph{28th {USENIX} Security Symposium ({USENIX} Security
  19)}.\hskip 1em plus 0.5em minus 0.4em\relax Santa Clara, CA: {USENIX}
  Association, Aug. 2019, pp. 801--818. [Online]. Available:
  \url{https://www.usenix.org/conference/usenixsecurity19/presentation/das}
\BIBentrySTDinterwordspacing

\bibitem{DataAvailability}
EthHub, ``Data availability,''
  \url{https://ethereum.org/en/developers/docs/data-availability}, accessed on
  05/21/2023.

\bibitem{eip-4488}
\BIBentryALTinterwordspacing
V.~Buterin and A.~Dietrichs, ``Eip-4488: Transaction calldata gas cost
  reduction with total calldata limit,''
  \url{https://eips.ethereum.org/EIPS/eip-4488}, Nov 2021. [Online]. Available:
  \url{https://eips.ethereum.org/EIPS/eip-4488}
\BIBentrySTDinterwordspacing

\bibitem{zkrollup}
EthHub, ``Zk-rollups,''
  \url{https://docs.ethhub.io/ethereum-roadmap/layer-2-scaling/zk-rollups/},
  accessed on 07/13/2022.

\bibitem{oprollup}
------, ``Optimistic rollups,''
  \url{https://docs.ethhub.io/ethereum-roadmap/layer-2-scaling/optimistic\_rollups/},
  accessed on 07/13/2022.

\bibitem{eip-4844}
\BIBentryALTinterwordspacing
V.~Buterin, D.~L. Dankrad~Feist, G.~Kadianakis, M.~Garnett, and A.~Dietrichs,
  ``Eip-4844: Shard blob transactions,''
  \url{https://eips.ethereum.org/EIPS/eip-4844}, Feb 2022. [Online]. Available:
  \url{https://eips.ethereum.org/EIPS/eip-4844}
\BIBentrySTDinterwordspacing

\bibitem{CancunUpgrade}
Ethereum, ``Cancun network upgrade specification,''
  \url{https://github.com/ethereum/execution-specs/blob/master/network-upgrades/mainnet-upgrades/cancun.md#included-eips},
  accessed on 05/21/2023.

\bibitem{qin2022quantifying}
K.~Qin, L.~Zhou, and A.~Gervais, ``Quantifying blockchain extractable value:
  How dark is the forest?'' in \emph{2022 IEEE Symposium on Security and
  Privacy (SP)}.\hskip 1em plus 0.5em minus 0.4em\relax IEEE, 2022, pp.
  198--214.

\bibitem{Ekiden:2019}
R.~Cheng, F.~Zhang, J.~Kos, W.~He, N.~Hynes, N.~Johnson, A.~Juels, A.~Miller,
  and D.~Song, ``{Ekiden: A Platform for Confidentiality-Preserving,
  Trustworthy, and Performant Smart Contracts},'' \emph{2019 IEEE European
  Symposium on Security and Privacy (EuroS\&P)}, vol.~00, pp. 185--200, 2019.

\bibitem{CONFIDE:SIGMOD20}
D.~Maier, R.~Pottinger, A.~Doan, W.-C. Tan, A.~Alawini, H.~Q. Ngo, Y.~Yan,
  C.~Wei, X.~Guo, X.~Lu, X.~Zheng, Q.~Liu, C.~Zhou, X.~Song, B.~Zhao, H.~Zhang,
  and G.~Jiang, ``{Confidentiality Support over Financial Grade Consortium
  Blockchain},'' 2020, pp. 2227--2240.

\bibitem{FrassettoNDSS22POSE}
T.~Frassetto, P.~Jauernig, D.~Koisser, D.~Kretzler, B.~Schlosser, S.~Faust, and
  A.-R. Sadeghi, ``Pose: Practical off-chain smart contract execution,'' in
  \emph{Proceedings of the 2022 Network and Distributed System Security
  Symposium}, vol. abs/2210.07110, 2022.

\bibitem{Choudhuri'17-FairMPC}
A.~R. Choudhuri, M.~Green, A.~Jain, G.~Kaptchuk, and I.~Miers, ``{Fairness in
  an Unfair World: Fair Multiparty Computation from Public Bulletin Boards},''
  ser. Proceedings of the 2017 ACM SIGSAC Conference on Computer and
  Communications Security, 2017, pp. 719--728.

\bibitem{ZEXE:SP20}
S.~Bowe, A.~Chiesa, M.~Green, I.~Miers, P.~Mishra, and H.~Wu, ``{ZEXE: Enabling
  Decentralized Private Computation},'' \emph{2020 IEEE Symposium on Security
  and Privacy}, 2020.

\bibitem{secondstate'20}
S.~State and O.~Labs, ``{Confidential Ethereum Smart Contracts},'' Tech. Rep.,
  12 2020.

\bibitem{ccf2019}
M.~Russinovich, E.~Ashton, C.~Avanessians, M.~Castro, A.~Chamayou, S.~Clebsch,
  and et~al., ``Ccf: A framework for building confidential verifiable
  replicated services,'' Microsoft Research and Microsoft Azure, Tech. Rep.,
  Apr. 2019.

\bibitem{Cavallaro2019Tesseract}
L.~Cavallaro, J.~Kinder, X.~Wang, J.~Katz, I.~Bentov, Y.~Ji, F.~Zhang,
  L.~Breidenbach, P.~Daian, and A.~Juels, ``{Tesseract: Real-Time
  Cryptocurrency Exchange Using Trusted Hardware},'' ser. Proceedings of the
  2019 ACM SIGSAC Conference on Computer and Communications Security, 2019, pp.
  1521--1538.

\bibitem{MPCpenalties'16}
E.~Weippl, S.~Katzenbeisser, C.~Kruegel, A.~Myers, S.~Halevi, R.~Kumaresan, and
  I.~Bentov, ``{Amortizing Secure Computation with Penalties},''
  \emph{Proceedings of the 2016 ACM SIGSAC Conference on Computer and
  Communications Security}, pp. 418--429, 2016.

\bibitem{MPCBitcoin'16}
E.~Weippl, S.~Katzenbeisser, C.~Kruegel, A.~Myers, S.~Halevi, R.~Kumaresan,
  V.~Vaikuntanathan, and P.~N. Vasudevan, ``{Improvements to Secure Computation
  with Penalties},'' \emph{Proceedings of the 2016 ACM SIGSAC Conference on
  Computer and Communications Security}, pp. 406--417, 2016.

\bibitem{PokerBitcoin'15}
I.~Ray, N.~Li, C.~Kruegel, R.~Kumaresan, T.~Moran, and I.~Bentov, ``How to use
  bitcoin to play decentralized poker,'' \emph{Proceedings of the 22nd ACM
  SIGSAC Conference on Computer and Communications Security}, pp. 195--206,
  2015.

\bibitem{FoteiniASIACRYPT20}
\BIBentryALTinterwordspacing
F.~Baldimtsi, A.~Kiayias, T.~Zacharias, and B.~Zhang, ``Crowd verifiable
  zero-knowledge and end-to-end verifiable multiparty computation,'' in
  \emph{Advances in Cryptology – ASIACRYPT 2020: 26th International
  Conference on the Theory and Application of Cryptology and Information
  Security, Daejeon, South Korea, December 7–11, 2020, Proceedings, Part
  III}.\hskip 1em plus 0.5em minus 0.4em\relax Berlin, Heidelberg:
  Springer-Verlag, 2020, p. 717–748. [Online]. Available:
  \url{https://doi.org/10.1007/978-3-030-64840-4_24}
\BIBentrySTDinterwordspacing

\bibitem{OzdemirSecurity22}
\BIBentryALTinterwordspacing
A.~Ozdemir and D.~Boneh, ``Experimenting with collaborative {zk-SNARKs}:
  {Zero-Knowledge} proofs for distributed secrets,'' in \emph{31st
  $\{$USENIX$\}$ Security Symposium ($\{$USENIX$\}$ Security 22)}.\hskip 1em
  plus 0.5em minus 0.4em\relax Boston, MA: USENIX Association, Aug. 2022, pp.
  4291--4308. [Online]. Available:
  \url{https://www.usenix.org/conference/usenixsecurity22/presentation/ozdemir}
\BIBentrySTDinterwordspacing

\bibitem{solc}
\BIBentryALTinterwordspacing
Ethereum, ``Solc 0.8.10,''
  \url{https://github.com/ethereum/solidity/releases/tag/v0.8.10}, July 2021.
  [Online]. Available:
  \url{https://github.com/ethereum/solidity/releases/tag/v0.8.10}
\BIBentrySTDinterwordspacing

\bibitem{costan2016intel}
V.~Costan and S.~Devadas, ``Intel sgx explained.'' \emph{IACR Cryptol. ePrint
  Arch.}, vol. 2016, no.~86, pp. 1--118, 2016.

\bibitem{evm}
\BIBentryALTinterwordspacing
E.~Foundation, ``Ethereum virtual machine,'' Dec 2020. [Online]. Available:
  \url{https://ethereum.org/en/developers/docs/evm/}
\BIBentrySTDinterwordspacing

\bibitem{gruss2017strong}
D.~Gruss, J.~Lettner, F.~Schuster, O.~Ohrimenko, I.~Haller, and M.~Costa,
  ``Strong and efficient cache side-channel protection using hardware
  transactional memory,'' in \emph{26th $\{$USENIX$\}$ Security Symposium
  ($\{$USENIX$\}$ Security 17)}, 2017, pp. 217--233.

\bibitem{shih2017t}
M.-W. Shih, S.~Lee, T.~Kim, and M.~Peinado, ``T-sgx: Eradicating
  controlled-channel attacks against enclave programs.'' in \emph{NDSS}, 2017.

\bibitem{Lang2022MoLEMO}
F.~Lang, W.~Wang, L.~Meng, J.~Lin, Q.~Wang, and L.~Lu, ``Mole: Mitigation of
  side-channel attacks against sgx via dynamic data location escape,''
  \emph{Proceedings of the 38th Annual Computer Security Applications
  Conference}, 2022.

\bibitem{noorman2013sancus}
J.~Noorman, P.~Agten, W.~Daniels, R.~Strackx, A.~Van~Herrewege, C.~Huygens,
  B.~Preneel, I.~Verbauwhede, and F.~Piessens, ``Sancus: Low-cost trustworthy
  extensible networked devices with a zero-software trusted computing base,''
  in \emph{22nd $\{$USENIX$\}$ Security Symposium ($\{$USENIX$\}$ Security
  13)}, 2013, pp. 479--498.

\bibitem{costan2016sanctum}
V.~Costan, I.~Lebedev, and S.~Devadas, ``Sanctum: Minimal hardware extensions
  for strong software isolation,'' in \emph{25th $\{$USENIX$\}$ Security
  Symposium ($\{$USENIX$\}$ Security 16)}, 2016, pp. 857--874.

\bibitem{Biondo2018codereuse}
\BIBentryALTinterwordspacing
A.~Biondo, M.~Conti, L.~Davi, T.~Frassetto, and A.-R. Sadeghi, ``The
  guard{\textquoteright}s dilemma: Efficient {Code-Reuse} attacks against intel
  {SGX},'' in \emph{27th USENIX Security Symposium (USENIX Security 18)}.\hskip
  1em plus 0.5em minus 0.4em\relax Baltimore, MD: USENIX Association, Aug.
  2018, pp. 1213--1227. [Online]. Available:
  \url{https://www.usenix.org/conference/usenixsecurity18/presentation/biondo}
\BIBentrySTDinterwordspacing

\bibitem{Brasser2017cacheattack}
\BIBentryALTinterwordspacing
F.~Brasser, U.~M{\"u}ller, A.~Dmitrienko, K.~Kostiainen, S.~Capkun, and A.-R.
  Sadeghi, ``Software grand exposure: {SGX} cache attacks are practical,'' in
  \emph{11th USENIX Workshop on Offensive Technologies (WOOT 17)}.\hskip 1em
  plus 0.5em minus 0.4em\relax Vancouver, BC: USENIX Association, Aug. 2017.
  [Online]. Available:
  \url{https://www.usenix.org/conference/woot17/workshop-program/presentation/brasser}
\BIBentrySTDinterwordspacing

\bibitem{Bulck2018foreshadow}
\BIBentryALTinterwordspacing
J.~V. Bulck, M.~Minkin, O.~Weisse, D.~Genkin, B.~Kasikci, F.~Piessens,
  M.~Silberstein, T.~F. Wenisch, Y.~Yarom, and R.~Strackx, ``Foreshadow:
  Extracting the keys to the intel {SGX} kingdom with transient {Out-of-Order}
  execution,'' in \emph{27th USENIX Security Symposium (USENIX Security
  18)}.\hskip 1em plus 0.5em minus 0.4em\relax Baltimore, MD: USENIX
  Association, Aug. 2018, p. 991{\textendash}1008. [Online]. Available:
  \url{https://www.usenix.org/conference/usenixsecurity18/presentation/bulck}
\BIBentrySTDinterwordspacing

\bibitem{l1tf}
\BIBentryALTinterwordspacing
Intel, ``Resources and response to side channel l1 terminal fault,'' Dec 2021.
  [Online]. Available:
  \url{https://www.intel.com/content/www/us/en/architecture-and-technology/l1tf.html?wapkw=l1tf}
\BIBentrySTDinterwordspacing

\bibitem{sgxfail}
``How stuff gets exposed,'' \url{https://sgx.fail/}, Jan 2022.

\end{thebibliography}

\clearpage

\label{sec:appendix}


\begin{strip}
\centering
\Large{Supplementary Material for \textit{``\codename: Enable Secure and Cheap \acrlong{mpt}s on Legacy Blockchains by a Minimally Trusted \acrshort{tee} Network''}}
\vspace{0.5cm}
\end{strip}

\subsection{\acrshort{mpt}-enabled contracts}\label{sec:model}
We have introduced the program $\mathcal{F}$, verifier $\mathcal{V}$, and enclave for achieving an \acrshort{mpt}. Here we model the privacy policy $\mathcal{P}$ we used for better managing parties' private data on-chain and specify the privacy demand of the \acrshort{mpt}.

Since each $\mathcal{E}$ need to encrypt/decrypt states before evaluating $\mathcal{F}$, $\mathcal{E}$ must aware about the states sets to read and write sets of $\mathcal{F}$. Therefore, we bind a privacy policy $\mathcal{P}$ to each $\mathcal{F}$.
$$
\begin{array}{lll}
    adr_{\mathcal{V}} & := & \{0,1\}^{*} \\
    v & := & {[\mathrm{a}-\mathrm{zA}-\mathrm{z} 0-9]+} \\ 
    P & := & \{0,1\}^{*} \\ 
    \mathcal{P}_{v} &:= & \emptyset~|~(v: P)~|~\left(v: P^?\right) \\
    \mathcal{P}_{\mathcal{F}} &:= & \{ \\
    & & \qquad \mathcal{P}_x := \{ \mathcal{P}_{v}\}^* \\ 
    & & \qquad \mathcal{P}_s  := \{ \mathcal{P}_{v}\}^* \\ 
    & & \qquad\mathcal{P}_{s'} := \{ \mathcal{P}_{v}\}^* \\ 
    & & \qquad \mathcal{P}_r := \{ \mathcal{P}_{v}\}^*\\
    & & \} \\
\end{array}
$$

A privacy policy $\mathcal{P}$ is modeled as the above. $adr_{\mathcal{V}}, adr_{\mathcal{V}}$ denote the address of its corresponding deployed $\mathcal{V}$ and verifier contract $\mathcal{V}$ on $BC$, respectively. $v$ refers to the identifiers of variables. The $P$ refers to parties' addresses. ($\mathcal{P}_x$) refers to transaction parameters of $\mathcal{F}$. $\mathcal{P}_s$ refers to states variables to be read to evaluate $\mathcal{F}$. $\mathcal{P}_{s'}$ refers to states variables to update by the $\mathcal{F}$. $\mathcal{P}_r$ refers to return variables of $\mathcal{F}$. Each variable is denoted by the tuple $(v:P)$, containing its identifier $v$ and the address of of its owner $P$ (\ie, the party that the variable private to). $v$ is owned by $P$ meaning that $v$ is confidential to $P$. Consequently, $\mathcal{E}$ expect to receive the $v$ from $P$ and commit $v$ with $P$'s public key. 
If the owner of an variable is unknown before \acrshort{mpt}, we write $(n:P^?)$. The unknown party will be settled after the \textit{negotiation phase} in Section~\ref{sec:overview}.

\subsection{Notations and Definitions} \label{sec:security-definition}
In this section, we fine-tuned the notation system of~\cite{FastKitten'19, ren2022Cloak} to denote variables involved in \codename. 
\subsubsection{Common notations}
Generally, we denote a domain as $\mathbb{S}$ and its $n$-ary Cartesian power $\mathbb{S} \times \mathbb{S} \times \cdots \times \mathbb{S}$ as $\mathbb{S}^n$. Therefore, each $\textit{\textbf{s}} \in \mathbb{S}^n$ is a array $[ s_1, \cdots, s_n ]$ and we refer $\textit{\textbf{s}}[i]$ or $s_i$ to the $i$-th element of $\textit{\textbf{s}}$. Moreover, $\mathbb{S}^{n\times m}$ denotes the set of all $n$-ny-$m$ matrices consisting of elements from $\mathbb{S}$. Similarly, we denote $\textbf{S}[i][j]$ as the element in $i$-th row and $j$-th column of $\textbf{S}$, $\textbf{S}[i][\cdot]$ as the $i$-th row, and $\textbf{S}[\cdot][j]$ as the $j$-th column.

\subsubsection{Coins}
We define a set $\mathbb{D}_{coin}$ as a \emph{coin domain}, which includes all possible balance of parties' global coins and is a subset of non-negative rational numbers $\mathbb{Q}\geq 0$. Therefore, we define a \emph{coin array} $\textit{\textbf{q}}\in \mathbb{D}_{coin}^n$ where $Q_i$ denotes the balance of party $\textit{P}_i$'s global coins. 
Then, we define the set $\mathbb{D}_{dep}\gets\mathbb{D}_{coin}\backslash  \{0\}$ as a \emph{deposit domain}, and define a \emph{deposit array} $\textit{\textbf{d}}\in \mathbb{D}_{dep}^n$ where $\textit{\textbf{d}}[i]$ denotes the deposit of party $\textit{P}_i$ for joining an \acrshort{mpt}. 

\subsubsection{\acrlong{mpt}s}
We define a set $\mathbb{D}_{pa}$ as a \emph{plaintext domain} which is application-specific. Therefore, for each \acrshort{mpt}, we have its plaintext \emph{parameter array} $\textit{\textbf{x}}$, \emph{old state array} $\textit{\textbf{s}}$, \emph{new state array} $\textit{\textbf{s}}'$, and \emph{return array array} $\textit{\textbf{r}}$, where $\textit{\textbf{x}}, \textit{\textbf{s}}, \textit{\textbf{s}}', \textit{\textbf{r}} \in \mathbb{D}_{pa}^n$. Correspondingly, we define a set $\mathbb{D}_{cm}$ as a \emph{cryptography commitment domain} which is specific to the cryptography commitment algorithm we adopted in Section~\ref{sec:overview}. Then, for each \acrshort{mpt}, we denote its \emph{parameter commitment array}, \emph{old state commitment array}, \emph{new state commitment array}, and \emph{return value commitment array} as $\textit{\textbf{c}}_x, \textit{\textbf{c}}_s, \textit{\textbf{c}}_{s'}, \textit{\textbf{c}}_r$, respectively, where $\textit{\textbf{c}}_x, \textit{\textbf{c}}_s, \textit{\textbf{c}}_{s'}, \textit{\textbf{c}}_r \in \mathbb{D}_{cm}^n$.

We define a \emph{party domain} $\mathbb{D}_{addr}$. $\mathbb{D}_{addr}$ is the set of all possible addresses of parties, thus depends on the address generation algorithm the $BC$ adopted. Then, the parties of an \acrshort{mpt} are modeled as a \emph{party array} $\textit{\textbf{P}}$ where $P_i$ denotes the $i$-th party of $\textit{\textbf{P}}$ and $P_i\in \mathbb{D}_{addr} $. We define the \emph{target function} of an \acrshort{mpt} which multiple parties jointly evaluate as $\mathcal{F}$, and the \emph{privacy policy} of an \acrshort{mpt} as $\mathcal{P}$ which specifies the meta data of $\mathcal{F}$, \eg, expected $\textit{\textbf{x}}, \textit{\textbf{s}}, \textit{\textbf{s}}', \textit{\textbf{r}}$. Then, we denote $\mathcal{F}_{\mathcal{P}}$ as a $\mathcal{P}$-conformed $\mathcal{F}$.

\begin{algorithm}[!htbp]
    \caption{Evaluation function}
    \label{alg:eval}
    \LinesNumbered
    \small
    \KwIn{An $n$-party \acrshort{mpt} $\mathcal{F}$ and its policy ${\mathcal{P}}$, a parameter array $\textit{\textbf{x}}$, a parameter key array $\textit{\textbf{k}}_x$, a old state array $\textit{\textbf{s}}$, a a old state key array $\textit{\textbf{k}}_s$, a old state commitment array $\textit{\textbf{c}}_s$, and a party array $\textit{\textbf{P}}$.} 
    \KwOut{A new state array $\textit{\textbf{s}}'$, new state key array $\textit{\textbf{k}}_{s'}$, return value array $\textit{\textbf{r}}$, return value key array $\textit{\textbf{k}}_{r}$, new state commitment array $\textit{\textbf{c}}_{s'}$, return value commitment array $\textit{\textbf{c}}_{r}$, parameter commitment array $\textit{\textbf{c}}_{x}$, and a $proof$.} 
 
    \SetKwFunction{Fevaluate}{\FuncSty{eval}}
    \SetKwProg{Fn}{Function}{}{}
    \Fn{\Fevaluate{$\mathcal{F}, \mathcal{P}, \textit{\textbf{x}}, \textit{\textbf{k}}_x, \textit{\textbf{c}}_s, \textit{\textbf{P}}$}}{
        \textbf{foreach} $c_{s_i}$ \textbf{in} $\textit{\textbf{c}}_{s}$ \\
        \quad \textbf{assert} $c_{s_i} = [\FuncSty{Enc}_{k_{s_i}}(s_i), \FuncSty{Enc}_{k_{ie}}(k_{s_i}), P_i]$ \\
        $ \textit{\textbf{s}}', \textit{\textbf{r}} \gets \mathcal{F}_{\mathcal{P}}(\textit{\textbf{s}}, \textit{\textbf{x}})$ \\
        $\textit{\textbf{k}}_{s'}, \textit{\textbf{k}}_r \gets Gen(1^\kappa)$ \\
        $c_{s'_i} \gets [\FuncSty{Enc}_{k_{s'_i}}(s'_i), \FuncSty{Enc}_{k_{ie}}(k_{s'_i}), P_i]$ \\
        $c_{r_i} \gets [\FuncSty{Enc}_{k_{r_i}}(r_i), \FuncSty{Enc}_{k_{ie}}(k_{r_i}), P_i]$ \\
        $c_{x_i} \gets [\FuncSty{Enc}_{k_{x_i}}(x_i), \FuncSty{Enc}_{k_{ie}}(k_{x_i}), P_i]$ \\
        $proof\gets [H_{\mathcal{P}}, H_{\mathcal{F}}, H_{\textit{\textbf{c}}_s}] $ \\
        \Return $(\textit{\textbf{s}}', \textit{\textbf{k}}_{s'}, \textit{\textbf{r}}, \textit{\textbf{k}}_r, \textit{\textbf{c}}_{s'}, \textit{\textbf{c}}_{r}, \textit{\textbf{c}}_{x}, proof)$
    }
\end{algorithm}

\subsubsection{Protocol execution}
While $\textit{\textbf{P}}$, $\textit{\textbf{E}}$, and $\mathbfcal{E}$ denote the party array, executor array and \acrshort{tee} array of an \acrshort{mpt}, respectively, we define $\textit{\textbf{P}}_H$ and $\textit{\textbf{E}}_H$ as the honest parties in $\textit{\textbf{P}}$ and $\textit{\textbf{E}}$ respectively. $\textit{\textbf{P}}_M$ and $\textit{\textbf{E}}_M$ denote the malicious parties in $\textit{\textbf{P}}$ and malicious executors of \acrshort{tee}s, \ie, $\textit{\textbf{P}}_M\gets\textit{\textbf{P}}\backslash \textit{\textbf{P}}_H$, $\textit{\textbf{E}}_M\gets\textit{\textbf{E}}\backslash \textit{\textbf{E}}_H$. For convenience, we also define $\textit{\textbf{P}}^+\gets \textit{\textbf{P}}\cup \textit{\textbf{E}}$ and $\textit{\textbf{P}}^+_M\gets \textit{\textbf{P}}_M\cup \textit{\textbf{E}}_M$.

According to our adversary model in Section~\ref{sec:overview}, \codename protocol $\pi_{\codename}$, or simply $\pi$, proceeds in presence of an byzantine adversary $\mathcal{A}$ who can corrupts all-but-one $P_i\in \textit{\textbf{P}}^+$. And we define a \emph{coin balance array} $\textit{\textbf{Q}} \in \mathbb{D}^{n+m}_{coin}$. $Q_i|_{i<n}$ denotes the coin balance of $P_i\in \textit{\textbf{P}}$ pre-deposited to $k_{\mathbfcal{E}}$. $Q_{n+i}|_{i<m}$ denotes the coin pre-deposited balance of $\mathcal{E}_i\in \mathbfcal{E}$.

Classically, we define any \emph{protocol execution} of $\pi$ under the adversary $\mathcal{A}$ as $REAL_{\pi, \mathcal{A}}$. The inputs of an execution include an $n$-party \acrshort{mpt} $\mathcal{F}$ and its policy ${\mathcal{P}}$, a parameter array $\textit{\textbf{x}}$, a parameter key array $\textit{\textbf{k}}_x$, a old state array $\textit{\textbf{s}}$, a a old state key array $\textit{\textbf{k}}_s$, a old state commitment array $\textit{\textbf{c}}_s$, a party array $\textit{\textbf{P}}$, a deposit array $\textit{\textbf{q}}$ and a account coin balance array $\textit{\textbf{Q}}$. Therefore, we formalize a protocol execution as follows. 
\begin{equation*}
    \begin{aligned}
    \qquad& \textit{\textbf{Q}}', \textit{\textbf{s}}', \textit{\textbf{k}}_{s'}, \textit{\textbf{r}}, \textit{\textbf{k}}_r, \textit{\textbf{c}}_{s'}, \textit{\textbf{c}}_{r}, \textit{\textbf{c}}_{x}, proof, sta \\
    \qquad& \qquad \gets REAL_{\pi, \mathcal{A}}(\textit{\textbf{Q}}, \mathcal{F}, \mathcal{P}, \textit{\textbf{x}}, \textit{\textbf{k}}_x, \textit{\textbf{c}}_s, \textit{\textbf{P}}, \textit{\textbf{q}})
    \end{aligned}
\end{equation*}
The outputs of $\pi$ include a new coin balance array $\textit{\textbf{Q}}'$ after the execution, new state array $\textit{\textbf{s}}'$, new state key array $\textit{\textbf{k}}_{s'}$, return value array $\textit{\textbf{r}}$, return value key array $\textit{\textbf{k}}_r$, and the commitment array of new states, return values, and parameters, \ie, $\textit{\textbf{c}}_{s'}, \textit{\textbf{c}}_{r}, \textit{\textbf{c}}_{x}$, respectively, and $proof$ of the \acrshort{mpt}-caused state transition.
            
\subsubsection{Security goals}
We first define the basic \emph{correctness} property. Intuitively, \emph{correctness} states that if all entities in $\textit{\textbf{P}}^+$ behave honestly, $\forall P_i\in \textit{\textbf{P}}$ obtain their correct \acrshort{mpt} outputs correspondingly and collateral back. 

\begin{definition}[Correctness]
    For any $n$-party \acrshort{mpt} $\mathcal{F}_{\mathcal{P}}$, $\textit{\textbf{q}}\in \mathbb{D}^n_{dep}$, $\textit{\textbf{s}} \in \mathbb{D}^n_{pa}$, $\textit{\textbf{x}}\in \mathbb{D}^n_{pa}$ and $\textit{\textbf{Q}}\in\mathbb{D}^{n}_{coin}$, there is a negligible function $\epsilon$ that for the output of the protocol $REAL_{\pi}(\textit{\textbf{Q}}, \mathcal{F}, \mathcal{P}, \textit{\textbf{x}}, \textit{\textbf{k}}_x, \textit{\textbf{c}}_s, \textit{\textbf{P}}, \textit{\textbf{q}})$ and $\forall P_i \in \textit{\textbf{P}}$
    $$\left| Pr\left[
        \begin{array}{c}
        \makecell[l]{
        (\textit{\textbf{s}}', \textit{\textbf{k}}_{s'}, \textit{\textbf{r}}, \textit{\textbf{k}}_r, \textit{\textbf{c}}_{s'}, \textit{\textbf{c}}_{r}, \textit{\textbf{c}}_{x}, proof) \\
        \qquad = \FuncSty{eval}(\mathcal{F}, \mathcal{P}, \textit{\textbf{x}}, \textit{\textbf{k}}_x, \textit{\textbf{c}}_s, \textit{\textbf{P}})
        } \\
        Q'_i \geq Q_i \\
        sta = \code{COMPLETED} \\
        \end{array} \right]-1\right| \leq \epsilon$$
\end{definition}

\begin{definition}[Confidentiality]
    For any $n$-party \acrshort{mpt} $\mathcal{F}_{\mathcal{P}}$, any adversary $\mathcal{A}$ corrupting parties from $\textit{\textbf{P}}^+_M$ in which $\textit{\textbf{P}}_M \subsetneqq \textit{\textbf{P}}$, any $\textit{\textbf{q}}\in \mathbb{D}^n_{dep}$, $\textit{\textbf{s}} \in \mathbb{D}^n_{pa}$, $\textit{\textbf{x}} \in \mathbb{D}^n_{pa}$ and $\textit{\textbf{Q}}\in\mathbb{D}^{n}_{coin}$, the protocol $REAL_{\pi, \mathcal{A}}(\textit{\textbf{Q}}, \mathcal{F}, \mathcal{P}, \textit{\textbf{x}}, \textit{\textbf{k}}_x, \textit{\textbf{c}}_s, \textit{\textbf{P}}, \textit{\textbf{q}})$ is such that: There is a negligible function $\epsilon$ ensuring that $\forall x^*_1, s^*_1, s^{'*}_1, r^{*}_1, x^*_2, s^*_2, s^{'*}_2, r^{*}_2, \in \mathbb{D}_{pa}$ and $\forall P_i \in \textit{\textbf{P}}_{H}$~:
    
    $$
    \makecell[c]{
        \left| Pr[x_i, s_i = x^{*}_1, s^{*}_1] - Pr[x_i, s_i = x^{*}_2, s^{*}_2] \right| ~\leq~ \epsilon \\
        \text{and} \\
        \left| Pr[s'_i, r_i = s^{'*}_1, r^{*}_1] - Pr[s'_i, r_i = s^{'*}_2, r^{*}_2] \right| ~\leq~ \epsilon
        \\
    }
    $$
\end{definition}


\begin{definition}[Data availability]
    For any $n$-party \acrshort{mpt} $\mathcal{F}_{\mathcal{P}}$, any adversary $\mathcal{A}$ corrupting parties from $\textit{\textbf{P}}^+$, any $\textit{\textbf{q}}\in \mathbb{D}^n_{dep}$, $\textit{\textbf{s}} \in \mathbb{D}^n_{pa}$, $\textit{\textbf{x}} \in \mathbb{D}^n_{pa}$ and $\textit{\textbf{Q}}\in\mathbb{D}^{n}_{coin}$, the protocol $REAL_{\pi, \mathcal{A}}(\textit{\textbf{Q}}, \mathcal{F}, \mathcal{P}, \textit{\textbf{x}}, \textit{\textbf{k}}_x, \textit{\textbf{c}}_s, \textit{\textbf{P}}, \textit{\textbf{q}})$ is such that: There is a negligible function $\epsilon$ satisfies that if $sta =\code{COMPLETED}$, one of the following statements must be true.
    
    \vspace{0.1cm}
    \begin{math} 
    \begin{aligned}
        & \text{(i)}~\textit{\textbf{E}}_M \subsetneqq \textit{\textbf{E}}: \forall E_i \in \textit{\textbf{E}}_{H},~\textit{there is a polynomial function}~f_{\mathcal{E}_i}\\ 
        & \qquad that~s'_i = f_{\mathcal{E}_i}(sk_{\mathbfcal{E}}, P_i, c_{s'_i}) \\
        & \text{(ii)}~\textit{\textbf{E}}_M = \textit{\textbf{E}}~\&~\textit{\textbf{P}}_M \subsetneqq \textit{\textbf{P}}: \forall P_i \in \textit{\textbf{P}}_{H},~\textit{there is a polynomial} \\
        & \qquad function~f_{\mathcal{P}_i}~that~s'_i = f_{\mathcal{E}_i}(sk_{P_i}, ad_{\mathbfcal{E}}, c_{s'_i}) \\
    \end{aligned}
    \end{math}
\end{definition}


\begin{definition}[Financial fairness]
    \label{def:fairness}
    For any $n$-party \acrshort{mpt} $\mathcal{F}_{\mathcal{P}}$, any adversary $\mathcal{A}$ corrupting parties from $\textit{\textbf{P}}^+_{M} \subsetneqq \textit{\textbf{P}}^+$, any $\textit{\textbf{q}}\in \mathbb{D}^n_{dep}$, $\textit{\textbf{s}} \in \mathbb{D}^n_{pa}$, $r\in \mathbb{D}^n_{pa}$ and $\textit{\textbf{Q}}\in\mathbb{D}^{n}_{coin}$, the output of the protocol $REAL_{\pi, \mathcal{A}}(\textit{\textbf{Q}}, \mathcal{F}, \mathcal{P}, \textit{\textbf{x}}, \textit{\textbf{k}}_x, \textit{\textbf{c}}_s, \textit{\textbf{P}}, \textit{\textbf{q}})$ is such that one of the following statements must be true:

    \vspace{0.1cm}
    \begin{math} 
        \begin{aligned}
            & \text{(i)}~sta\in \{\code{NEGOFAILED}, \code{COMPLETED}\},~\forall P_i\in \textit{\textbf{P}}^+: Q'_i\geq Q_i\\
            & \text{(ii)}~sta=\code{ABORTED},~\forall P_i\in \textit{\textbf{P}}^+_{H}:~Q'_i\geq Q_i~\text{and} \\
            & \qquad {\sum_{j\in \textit{\textbf{P}}^+_{M}} Q'_j < \sum_{j\in \textit{\textbf{P}}^+_{M}} Q_j}\\
        \end{aligned}
    \end{math}
\end{definition}

\begin{definition}[Delivery fairness]
    \label{def:standard-fairness}
    For any $n$-party \acrshort{mpt} $\mathcal{F}_{\mathcal{P}}$, any adversary $\mathcal{A}$ corrupting parties from $\textit{\textbf{P}}^+_{M}$ in which $\textit{\textbf{E}}_{M} \subsetneqq \textit{\textbf{E}}$, any $\textit{\textbf{q}}\in \mathbb{D}^n_{dep}$, $\textit{\textbf{s}} \in \mathbb{D}^n_{pa}$, $r\in \mathbb{D}^n_{pa}$ and $\textit{\textbf{Q}}\in\mathbb{D}^{n}_{coin}$, there is a negligible function $\epsilon$ that for the output of the protocol $REAL_{\pi, \mathcal{A}}(\textit{\textbf{Q}}, \mathcal{F}, \mathcal{P}, \textit{\textbf{x}}, \textit{\textbf{k}}_x, \textit{\textbf{c}}_s, \textit{\textbf{P}}, \textit{\textbf{q}})$, one of the following statements must be true:
    \vspace{0.1cm}
    
    \begin{math} 
        \begin{aligned}
            & \text{(i)}~\textit{\textbf{s}}', \textit{\textbf{r}} = \emptyset, \emptyset \\
            & \text{(ii)}~\textit{\textbf{s}}', \textit{\textbf{r}}, \neq \emptyset,\emptyset,~and~\text{the following two hold simultaneously}: \\ 
            &\qquad \text{(a)}~\forall P_i\in \textit{\textbf{P}}_H: \left| t_{s_i} - t_{r_i} \right| \leq \Delta \\
            &\qquad \text{(b)}~\forall P_i, P_j\in \textit{\textbf{P}}_H:\left| t_{s_i} - t_{s_j} \right| \leq \Delta~\text{and}~\left| t_{r_i} - t_{r_j} \right| \leq \Delta \\
        \end{aligned}
    \end{math}
\end{definition}

\begin{definition}[Delivery atomicity]
    \label{def:standard-fairness}
    For any $n$-party \acrshort{mpt} $\mathcal{F}_{\mathcal{P}}$, any adversary $\mathcal{A}$ corrupting parties from $\textit{\textbf{P}}^+_{M}$ in which $\textit{\textbf{E}}_{M} \subsetneqq \textit{\textbf{E}}$, any $\textit{\textbf{q}}\in \mathbb{D}^n_{dep}$, $\textit{\textbf{s}} \in \mathbb{D}^n_{pa}$, $r\in \mathbb{D}^n_{pa}$ and $\textit{\textbf{Q}}\in\mathbb{D}^{n}_{coin}$, there is a negligible function $\epsilon$ that for the output of the protocol $REAL_{\pi, \mathcal{A}}(\textit{\textbf{Q}}, \mathcal{F}, \mathcal{P}, \textit{\textbf{x}}, \textit{\textbf{k}}_x, \textit{\textbf{c}}_s, \textit{\textbf{P}}, \textit{\textbf{q}})$, one of the following statements must be true:
    \vspace{0.1cm}
    
    \begin{math} 
        \begin{aligned}
            & \text{(i)}~sta\in \{\emptyset, \code{NEGOFAILED}, \code{ABORTED}\},~and~\textit{\textbf{s}}', \textit{\textbf{r}} = \emptyset, \emptyset \\
            & \text{(ii)}~sta=\code{COMPLETED},~and~\textit{\textbf{s}}',   \textit{\textbf{r}}, \neq \emptyset
        \end{aligned}
    \end{math}
\end{definition}

\subsection{Security Proof} \label{sec:security-proof}
In this section, we claim that the following theorem holds in the \codename protocol $\pi_{\codename}$.

\begin{theorem-box}
    \begin{theorem}[Formal statement]
        Assume a \acrshort{cma} secure signature scheme, a \acrshort{cca2} encryption scheme, a hash function that is collision-resistant, preimage and second-preimage resistant. a \acrshort{tee} emulating the \acrshort{tee} ideal functionality and a $BC$ emulating the $BC$ ideal functionality, $\pi_{\codename}$ holds \textbf{correctness}, \textbf{confidentiality}, \textbf{public verifiability}, \textbf{data availability}, \textbf{financial fairness}, \textbf{delivery fairness}, and \textbf{delivery atomicity}.
    \end{theorem}
\end{theorem-box}

\subsubsection{Proof of correctness}
Consider adversaries absent in $\pi_{\codename}$. The evaluation of an \acrshort{mpt} starts by the specified $\mathcal{E}^*$ receiving an \acrshort{mpt} proposal $p\gets (H_{\mathcal{F}},H_{\mathcal{P}}, q, h_{neg})$ and starting the \textit{negotiation phase}. $\mathcal{E}^*$ first deterministically generates an id $id_p$ of the proposal and broadcast the $id_p$ with the proposal to $P_i \in \textit{\textbf{P}}$. When $\mathcal{E}^*$s collects satisfied acknowledgement from $\textit{\textbf{P}}$, it broadcasts the settled $p'$. In the \textit{execution phase}, $\mathcal{E}^*$ collects the plaintext inputs $\textit{\textbf{in}}$ from $\textit{\textbf{P}}$ and read $s_i$ from $BC.\textit{\textbf{c}}_{s}$. Then, $\mathcal{E}^*$ obtains the \acrshort{mpt}'s outputs by 
$$\textit{\textbf{s}}', \textit{\textbf{k}}_{s'}, \textit{\textbf{r}}, \textit{\textbf{k}}_r, \textit{\textbf{c}}_{s'}, \textit{\textbf{c}}_{r}, \textit{\textbf{c}}_{x}, proof \gets \FuncSty{eval}(\mathcal{F}, \mathcal{P}, \textit{\textbf{x}}, \textit{\textbf{k}}_x, \textit{\textbf{c}}_s, \textit{\textbf{P}})$$
Then it moves to the \textit{delivery phase}. $\mathcal{E}^*$ releases a $TX_{cmt}$ to commit the outputs without publishing the symmetric key ciphertext. Upon the only one $TX_{cmt}$ is confirmed on $BC$, each $\mathcal{E}$ reads the $TX_{cmt}$ to obtain the shared symmetric keys $k_{s'_i}, k_{r_i}$. Then, each $\mathcal{E}$ encrypts the keys $k_{s'_i}, k_{r_i}$ with the $k_{ie}$ and broadcasts a $TX_{com}$ to both $\textit{\textbf{P}}$ and $BC$ immediately. As no $P_i \in \textit{\textbf{P}}^+$ is punished, we have $Q'_i\gets Q_i\geq Q_i$.

Since all protocol messages are sent in secure channels between $\textit{\textbf{P}}$ and $\mathcal{E}$s and we ignore the leakage caused by $\mathcal{F}$ and parties' voluntarily revealing, the \textit{confidentiality} is axiomatic. Therefore, we proves \textit{data availability}, \textit{financial fairness}, and \textit{delivery ($\Delta-$)fairness} in the following.

\subsubsection{Proof of data availability}
According to the Algorithm~\ref{alg:cloak-service}, when $sta=\code{COMPLETED}$, there must be $\textit{\textbf{c}}_s$ published on $BC$. Recall the data structure of $c_{s'_i} \gets [\FuncSty{Enc}_{k_{s'_i}}(s'_i),\FuncSty{Enc}_{k_{ie}}(k_{s'_i}),P_i]$, we construct a polynomial function in Algorithm~\ref{alg:availability-function}. With the function, any $\mathcal{E}\in \mathbfcal{E}$ or $P_i \in \textit{\textbf{P}}$ can construct the newest states of all completed \acrshort{mpt} independently. Therefore, the data availability holds.

\begin{algorithm}[!htbp]
    \caption{States construction function}
    \label{alg:availability-function}
    \LinesNumbered
    \small
 
    \SetKwFunction{FconstructStates}{\FuncSty{constructStates}}
    \SetKwProg{Fn}{Function}{}{}
    \Fn{\FconstructStates{$sk, pk, c_{s'_i}$}}{
        $k_ie \gets \FuncSty{ECDH}(sk, pk)$ \\
        $k_{s'_i} \gets \FuncSty{Dec}_{k_ie}(c_{s'_i}[1]) $ \\
        $s'_i \gets \FuncSty{Dec}_{k_{s'_i}}(c_{s'_i}[0]) $ \\
        \Return $s'_i$ 
    }
\end{algorithm}

\subsubsection{Proof of financial fairness}
Here we prove that in all possible $sta$, the financial fairness of $\pi_{\codename}$ holds. First, we consider the \emph{Negotiation phase}. Briefly, we prove that if the phase does not complete successfully then the proposal will have $sta=\code{NEGOFAILED}$ and $\forall P_i \in \textit{\textbf{P}}_{H}$ stays financially neutral.
\begin{lemma-box}
    \begin{lemma}
        \label{lemma:financial-empty}
        If there $\exists P_i\in \textit{\textbf{P}}_H$ stays at $sta=\code{NEGOFAILED}$, then the statement (i) of the financial fairness property holds.
    \end{lemma}
\end{lemma-box}
\emph{Proof:}
There is only one cases when an $P_i\in \textit{\textbf{P}}_{H}$ has $sta=\code{NEGOFAILED}$: 
\begin{itemize}[leftmargin=5mm, parsep=0.5mm, topsep=0.5mm, partopsep=0.5mm]
    \item (i) $TX_{fneg}$ is confirmed on $BC$ after $Proc_\text{nneg}$.
\end{itemize}
Specifically, this scenario happens when the collected $\textit{ack}$ from both on-chain and off-chain channels cannot satisfy the settlement condition of \acrshort{mpt} proposal or $\exists P_i \in \textit{\textbf{P}}$ holds that $\textit{\textbf{Q}}_i \leq q$. No matter what reasons cause the failure, we require $\forall P_i \in \textit{\textbf{P}}_H$ identifying the $sta$ of an \acrshort{mpt} by reading it from the $BC$. As we assume that the $BC$ emulates the ideal blockchain functionality which achieves ideal consistency and availability, $\forall P_i \in \textit{\textbf{P}}$ can access the consistent $BC$ view. Therefore, if a $TX_{fneg}$ is successfully confirmed on-chain. The result will be the unique result of the proposal ensured by \codename contract $\mathcal{V}$, and $\forall P_i \in \textit{\textbf{P}}_H$ will immediately identify that $sta=\code{NEGOFAILED}$. Then $Q'_i=Q_i$, \ie, $Q'_i\geq Q_i$ holds.

\begin{lemma-box}
    \begin{lemma}
        \label{lemma:financial-completed}
        If $~\exists P_i\in \textit{\textbf{P}}_H$ such that $sta = \code{COMPLETED}$, then the statement (i) of the financial fairness property holds.
    \end{lemma}
\end{lemma-box}

\emph{Proof:} According to Algorithm~\ref{alg:cloak-service} , the protocol outputs  $sta = \code{COMPLETED}$ iff a transaction $TX_{com}$ is contained on $BC$ before the $h_{cp}+\tau_{com}$-th block. Therefore, $\forall P_i \in \textit{\textbf{P}}^+$ the $Q'_i = Q_i \geq Q_i$ holds.

Next, we show that the financial fairness also holds even if an \acrshort{mpt} fails by $\code{ABORTED}$ after an successful \emph{Negotiation phase}.
\begin{lemma-box}
    \begin{lemma}
        \label{lemma:financial-aborted}
        If $~\exists P_i\in \textit{\textbf{P}}_H$ is such that $sta = \code{ABORTED}$, then the statement (ii) of the financial fairness property holds.
    \end{lemma}
\end{lemma-box}

\emph{Proof:} There are two cases when $\exists P_i\in \textit{\textbf{P}}_H$ outputs $\code{ABORTED}$:
\begin{itemize}[leftmargin=5mm, parsep=0.5mm, topsep=0.5mm, partopsep=0.5mm]
    \item (i) Before the $h_{cp}+\tau_{com}$-th block, $TX_{pnsP}(id_p, \textit{\textbf{P}}'_M)$ is published on $BC$.
    \item (ii) After the $h_{cp}+\tau_{com}$-th block, $TX_{pnsT}(id_p)$ is published on $BC$.
\end{itemize}

We first consider the case (i) where $\exists P_j\in \textit{\textbf{P}}'_{M}$ does not provide inputs $in_j$ after the negotiation succeeded. According to Algorithm~\ref{alg:cloak-enclave}, the $\mathcal{E}^*$ releases a transaction $TX_{pnsP}(id_p, \textit{\textbf{P}}_{M})$ iff $E^*$ calls the $\mathcal{E}^*.punishParties$ with a $\acrshort{pop}_{resP}$ which proves that $P_j\in\textit{\textbf{P}}_{M}|_{\textit{\textbf{P}}_{M}\neq \emptyset}$ did not provide their inputs even though they were challenged by a $TX_{chaP}$. The $TX_{pnsP}$ will deduct coins of $\forall P_i \in \textit{\textbf{P}}_{M}$ by the \acrshort{mpt}-specific collaterals $q$. In other word, for $\forall P_i \in \textit{\textbf{P}}_{M}$, it holds that $Q'_i =Q_i - q_i$. Since $Q_i>q_i$, which has been ensured by $Proc_\text{nneg}$, and $\textit{\textbf{P}}_{M}\neq \emptyset$, it holds that $\sum_{j\in \textit{\textbf{P}}_{M}} Q'_j < \sum_{j\in \textit{\textbf{P}}_{M}} Q_j$. Notably, no malicious party earned coins in this case. 

Second, we consider the case (ii) which indicates that $TX_{com}$ fails to be contained before the $h_{cp}+\tau_{com}$-th block. Since the case (i) not happens, then either $\mathcal{E}^*$ have collected correct inputs from all parties, which means that $\textit{\textbf{P}}_M=\emptyset$, or $E^*$ detains the $TX_{pnsT}$ or $TX_{cmt}$, or $TX_{cmt}$ fails on validation, \eg, the old state commitments $\textit{\textbf{c}}_s$ that $\mathcal{E}^*$ read from and executed \acrshort{mpt} on has been changed, which fails the $verify(proof, H_{\mathcal{F}}, H_{\mathcal{P}}, H_{\textit{\textbf{c}}_s})$ in $TX_{cmt}$. 
In any case, when the timeout transaction $TX_{pnsT}$ is posted by an honest party on the $BC$, it $p'$ will be marked as $\code{ABORTED}$ and $\forall P_i\in \textit{\textbf{P}}$ gets \ie, $Q'_i = Q_i$. The $Q'_i \geq Q_i$ holds.

\begin{lemma-box}
    \begin{lemma}
        \label{lemma:financial-state}
        When $\pi_{\codename}$ terminates, it must hold $sta \in \{\code{NEGOFAILED}, \code{NEGOFAILED}, \code{COMPLETED}\}$.
    \end{lemma}
\end{lemma-box}
\emph{Proof:} As we stressed, $\forall P_i \in \textit{\textbf{P}}_H$ and $\forall E\in \textit{\textbf{E}}_H, \mathcal{E} \in \mathbfcal{E}$ identify current $sta$ from the $\mathcal{V}$ on $BC$. If an \acrshort{mpt} succeeds, a $TX_{com}$ must be sent, which leads to $std\gets \code{COMPLETED}$. Otherwise, we claim that there must be $std\gets \code{NEGOFAILED}/\code{NEGOFAILED}$. According to the Algorithm~\ref{alg:cloak-service}, there are additionally one temporary status. When $TX_{cmt}$ is accepted, it indicates that the \acrshort{mpt} outputs are successfully validated. Recall that $BC$ can continuously serve new transactions, $TX_{com}$ has no output validation logic, and at least one executor is honest. There must be a executor who can send $TX_{com}$ to set $sta \gets \code{COMPLETED}$.

\subsubsection{Delivery ($\Delta$-)fairness}
Recall that the Lemma~\ref{lemma:financial-state} holds. In the following, we prove that the  \textit{delivery ($\Delta$-)fairness} holds in all three values of $sta$  that $\pi_{\codename}$ terminates at.
We first consider the \textit{negotiation phase}. Intuitively, if no sufficient acknowledgement is collected, $\mathcal{E}^*$ cannot move to the \textit{Execution phase}, therefore no outputs are obtained or delivered.

\begin{lemma-box}
    \begin{lemma}
        If there exist an honest party $P_i$ staying at $sta =\code{NEGOFAILED}$, then the statement (i) of the delivery ($\Delta-$)fairness holds.
    \end{lemma}
\end{lemma-box}

\emph{Proof:} As proved in Lemma~\ref{lemma:financial-empty}, an honest party $P_i$ stays at $sta=\code{NEGOFAILED}$ only when there is a $TX_{fneg}$ being successfully confirmed on the $BC$. Consequently, the $\mathcal{E}^*$ with the \textit{Execution phase}. Therefore, parties in $\textit{\textbf{P}}$ obtain no outputs, \ie, $\textit{\textbf{s}}', \textit{\textbf{r}} = \emptyset, \emptyset$. 

\begin{lemma-box}
    \begin{lemma}
        If there exist an honest party $P_i$ such that $sta = \code{ABORTED}$, then the statement (i) of the delivery ($\Delta-$)fairness holds.
    \end{lemma}
\end{lemma-box}

\emph{Proof:} One of $\mathcal{E}$ releases the $TX_{com}$ only when it validates that the 
predecessor $TX_{cmt}$ has been confirmed on $BC$. When $sta=\code{ABORTED}$, it means that, according to Algorithm~\ref{alg:cloak-service}, the protocol terminates and there is no possibility for $sta=\code{COMMITTED}$, so as to releasing $TX_{com}$. Therefore, it holds that $\textit{\textbf{s}}', \textit{\textbf{r}} = \emptyset, \emptyset$. 

\begin{lemma-box}
    \begin{lemma}
        If there exist an honest party $P_i$ such that $sta = \code{COMPLETED}$, then the statement (ii) of the delivery ($\Delta-$)fairness holds.
    \end{lemma}
\end{lemma-box}

\emph{Proof:} According to Algorithm~\ref{alg:cloak-service}, $sta=\code{COMPLETED}$ only when $TX_{com}$ is accepted and confirmed by $BC$, which means that $TX_{com}$ is released by at least one $\mathcal{E}$s. In fact, if $TX_{cmt}$ has been confirmed on $BC$, any $\mathcal{E}\in\mathbfcal{E}$ can validate the $\acrshort{pop}_{cmt}$ of $TX_{cmt}$ and read the $\textit{\textbf{k}}_{s'}, \textit{\textbf{k}}_{r}$ from $TX_{cmt}$ to constructs and releases a $TK_{com}$. As we assume that $BC$ is ideally accessible to any honest entity. Therefore, say $TX_{cmt}$ is confirmed on $BC$ in a wall-time $t_{com}$, then the time of all honest entities in $\textit{\textbf{P}}^+$ knowing that $TX_{cmt}$ has been confirmed is also $t_{com}$, \ie, $t_i \gets t_{com}|_{t_i \in \textit{\textbf{t}}^+_{com}}$. Moreover, as $P_i \in \textit{\textbf{P}}_H$ undisturbedly obtain $TX_{com}$ from honest $E$s within the network latency $\Delta$, then we conclude that $\textit{\textbf{t}}_s=\textit{\textbf{t}}_r$, \ie, the (a) and (b) of (ii) are satisfied, if at least one honest $E$ exists.

\vfill

\end{document}